\documentclass[aps,preprint]{revtex4}
\usepackage{amsfonts}
\usepackage{amsmath}
\usepackage{amssymb}
\usepackage[colorlinks,linkcolor=blue,citecolor=blue,urlcolor=blue]{hyperref}
\usepackage{subfigure}
\usepackage{graphicx}
\usepackage{cleveref}
\usepackage{listings}
\usepackage{xcolor}
\usepackage{array}
\usepackage{url}

\providecommand{\U}[1]{\protect\rule{.1in}{.1in}}

\begin{document}

\preprint{CTP-SCU/2020027}
\title{Extended Phase Space Thermodynamics for Dyonic Black Holes with a
Power Maxwell Field}
\author{Feiyu Yao$^{a}$}
\email{yaofeiyu@stu.scu.edu.cn}
\author{Jun Tao$^{a}$}
\email{taojun@scu.edu.cn}
\affiliation{$^{a}$Center for Theoretical Physics, College of Physics, Sichuan
University, Chengdu, 610064, China}

\begin{abstract}
In this paper, we investigate the thermodynamics of dyonic black holes with
the presence of power Maxwell electromagnetic field in the extended phase
space, which includes the cosmological constant $\Lambda$ as a thermodynamic
variable. For a generic power Maxwell black hole with the electric charge
and magnetic charge, the equation of state is given as the function of
rescaled temperature $\tilde{T}$ in terms of other rescaled variables $%
\tilde{r}_{+}$, $\tilde{q}$ and $\tilde{h}$, where $r_{+}$ is the horizon
radius, $q$ is the electric charge and $h$ is some magnetic parameter. For
some values of $\tilde{q}$ and $\tilde{h}$, the phase structure of the black
hole is uniquely determined. Moreover the peculiarity of multiple
temperature with some fixed parameter configurations results in more rich
phase structures. Focusing on the power Maxwell Lagrangian with $\mathcal{L}%
\left( s\right) =s^{2}$, we obtain the corresponding phase diagrams in the $%
\tilde{q}$-$\tilde{h}$ plane, then analyse the black holes phase structure
and critical behaviour. For this case, the critical line is semi-infinite
and extends to $\tilde{h}=\infty$. We also examine thermal stabilities of
these black holes.
\end{abstract}

\keywords{}
\maketitle
\tableofcontents

\section{Introduction}

Black holes are intriguing concepts from the two cornerstones of modern
theoretical physics: General Relativity and Quantum Field Theory.
Classically black holes absorbed all matter and emitted nothing.
Superficially they had neither temperature nor entropy, and were
characterized by only a few basic parameters: mass, angular momentum, and
charge (if any) \cite{Israel:1967wq}. However all of this were changed by
the advent of quantum field theory in curved spacetime. The first indication
linking black holes and thermodynamics came from Hawking's area theorem \cite%
{Hawking:1971tu}, which states that the area of the event horizon of a black
hole can never decrease. Bekenstein subsequently noticed the resemblance
between this area law and the second law of thermodynamics \cite%
{Bekenstein:1972tm}, proposing that each black hole should be assigned an
entropy proportional to the area of its event horizon \cite%
{Bekenstein:1973ur}. Analogous to the laws of thermodynamics, Bardeen,
Carter and Hawking soon established the four laws of black hole mechanics 
\cite{Bardeen:1973gs}, where the surface gravity corresponds to the
temperature. Since Hawking discovered that black holes do emit radiation
with a blackbody spectrum \cite{Hawking:1975iha}, the idea of black hole
thermodynamics has convinced most physicists. And over the past four
decades, a preponderance of evidence suggested that it is a meaningful
subject.

According the sign of cosmological constant $\Lambda$, black holes can be
classified into asymptotically de Sitter (dS) black holes ($\Lambda >0$),
asymptotically Anti-de Sitter (AdS) black holes ($\Lambda < 0$) and
asymptotically flat black holes ($\Lambda$=0). Sufficiently large
asymptotically AdS (as compared to the AdS radius $l$) black holes, unlike
asymptotically flat black holes, have positive specific heat and can be in
stable equilibrium at a fixed temperature \cite{Hawking:1982dh}. Moreover
asymptotically AdS black holes, unlike asymptotically dS BH \cite%
{Gibbons:1977mu}, only have one horizon and one can define a good notion of
the asymptotic mass. So the asymptotically AdS black holes are always
popular subjects. Studying the phase transitions of asymptotically AdS black
holes is primarily motivated by AdS/CFT correspondence \cite%
{Maldacena:1997re}. Hawking and Page showed that a first-order phase
transition occurs between Schwarzschild AdS black holes and thermal AdS
space \cite{Hawking:1982dh}, which was later understood as a
confinement/deconfinement phase transition in the context of the AdS/CFT
correspondence \cite{Witten:1998zw}. For Reissner-Nordstrom (RN) AdS black
holes, authors of \cite{Chamblin:1999tk,Chamblin:1999hg} showed that their
critical behavior is similar to that of a Van der Waals liquid gas phase
transition.

Later, the asymptotically AdS black holes were studied in the context of
extended phase space thermodynamics, where the cosmological constant is
interpreted as thermodynamic pressure \cite{Dolan:2011xt,Kubiznak:2012wp}.
In this case, the black hole mass should be understood as enthalpy instead
of the internal energy; the first law was modified \cite{Kastor:2009wy}. The 
$P$-$V$ criticality study has been explored for various AdS black holes \cite%
{Wei:2012ui,Cai:2013qga,Xu:2014kwa,Frassino:2014pha,Dehghani:2014caa,Hennigar:2015esa}%
. It showed that the $P$-$V$ critical behaviors of AdS black holes are
similar to that of a Van der Waals liquid gas system.

Nonlinear electrodynamics (NLED) is an effective model incorporating quantum
corrections to Maxwell electromagnetic theory. Coupling NLED to gravity,
various NLED charged black holes were derived and discussed in a number of
papers \ \cite%
{Miskovic:2010ey,Miskovic:2010ui,Soleng:1995kn,AyonBeato:1998ub,Maeda:2008ha,Hendi:2017mgb,Tao:2017fsy,Guo:2017bru,Mu:2017usw,Wang:2018hwg}%
. The thermodynamics of generic NLED black holes in the extended phase space
have been considered in \cite%
{Wang:2018xdz,Bai:2020ieh,Wang:2019dzl,Gan:2018utc,Wang:2019kxp}. And
various particular NLED black holes were also considered, e.g., Born-Infeld
AdS black holes \cite{Fernando:2003tz,Zou:2013owa},\ power Maxwell invariant
black holes\ \cite{Hendi:2012um,Mo:2016jqd,Yerra:2018mni}, nonlinear
magnetic-charged dS black holes \cite{Nam:2018tpf}.

Among the various NLED, straightforward generalization of Maxwell's theory
leads to the so called power Maxwell (PM) theory described by a Lagrangian
density of the form $\mathcal{L}\left( s\right) =s^{p}$, where $s$ is the
Maxwell invariant, and $p$ is an arbitrary rational number. Clearly the
special value $p=1$ corresponds to linear electrodynamics. The solutions of
PM charged black holes and their interesting thermodynamics and geometric
properties have been examined in \cite%
{Hassaine:2007py,Hendi:2009zzb,Hendi:2009zza,Hassaine:2008pw,Maeda:2008ha,Hendi:2010zza,Hendi:2010bk,Hendi:2010zz,Hendi:2010kv}%
.

The PM theory is a toy model to generalize Maxwell theory which reduces to
it for $p=1$. One of the most important properties of the PM model in $d$%
-dimensions occurs when $p=d/4,$ where the PM theory becomes conformally
invariant and the trace of energy-momentum tensor vanishes, such as Maxwell
theory in four-dimensions. On the other hand, taking into account the
applications of the AdS/CFT correspondence to superconductivity, it has been
shown that the PM theory makes crucial effects on the condensation as well
as the critical temperature of the superconductor and its energy gap \cite%
{Jing:2011vz}.

A substantial gap in these studies is the absence of dyonic solution.
Authors of \cite{Bronnikov:2017xrt} derived a scheme of finding dyonic
solution in NLED coupled to GR by quadratures for an arbitrary Lagrangian
function $L\left( s\right) $ and a dyonic solution for the truncated
Born-Infeld theory. However there are still not many papers devoted to
studies of specific cases with dyonic solution.

In this paper, We first investigate the thermodynamic behavior of\ generic $%
d $-dimensional dyonic PM black holes in the extended phase space. Then, we
turn to study the phase structure and critical behavior of $8$-dimensional
dyonic PM black holes with an power exponent of $2$ by studying the phase
diagrams in the $q/l^{4.5}$-$h/l^{1.5}$ plane. After this Introduction, we
derive $d$-dimensional dyonic PM black hole solutions and discuss
thermodynamic properties of the black hole in section \ref{Sec:NBH}. In
section \ref{Sec:PMI2}, we study the phase structure and critical behavior
of 8-dimensional dyonic PM AdS black holes with an power exponent of $2$.
The phase diagram for the black hole in the $q/l^{4.5}$-$h/l^{1.5}$ plane is
given in FIG.\ \ref{fig:Re}, from which one can read the black hole's phase
structure and critical behavior. We also explore thermal stabilities of
these black holes. We summarize our results in section \ref{Sec:Con}. We
will use the units $\hbar=c=16\pi G=1$ for simplicity.

\section{Dyonic PM AdS Black Hole}

\label{Sec:NBH}

In this section, we derive the $d$-dimensional dyonic PM asymptotically AdS
black hole solution in the Einstein gravity and verify the thermodynamic
properties of the black hole. We first consider a $d$-dimensional model of
gravity coupled to a PM nonlinear electromagnetic field with action given by%
\begin{equation}
S_{\text{Bulk}}=\int d^{d}x\sqrt{-g}\left[ R-2\Lambda+\mathcal{L}\left(
s\right) \right] ,  \label{eq:Action}
\end{equation}
where 
\begin{equation}
\Lambda=-\frac{\left( d-1\right) \left( d-2\right) }{2l^{2}}
\end{equation}
is cosmological constant,%
\begin{equation}
s=\frac{1}{4}F_{\mu\nu}F^{\mu\nu}  \label{eq:maxwell}
\end{equation}
is the maxwell invarient, $F=dA=\partial_{\mu}A_{\nu}-\partial_{\nu}A_{\mu}$
and $A_{\mu}$ is the gauge potential. In our case the Lagrangian density has
the following form%
\begin{equation}
\mathcal{L}\left( s\right) =s^{p}.
\end{equation}
Taking the variation of the action $\left( \ref{eq:Action}\right) $ with
respect\ to $g_{\mu\nu}$ and $A_{\mu}$, one can get the equations of\
motion, they are%
\begin{align}
R_{\mu\nu}-\frac{1}{2}Rg_{\mu\nu}-\frac{\left( d-2\right) \left( d-1\right) 
}{2l^{2}}g_{\mu\nu} & =\frac{T_{\mu\nu}}{2}\text{,} \\
\partial_{\mu}\left( \sqrt{-g}G^{\mu\nu}\right) & =0\text{,}
\end{align}
where%
\begin{equation}
T_{\mu\nu}=g_{\mu\nu}\mathcal{L}\left( s\right) \mathbb{+}\frac {\partial%
\mathcal{L}\left( s\right) }{\partial s}F_{\mu}^{\text{ }\rho }F_{\nu\rho}
\end{equation}
is energy-momentum tensor and we defined the auxiliary field%
\begin{equation}
G^{\mu\nu}=\frac{\partial\mathcal{L}\left( s\right) }{\partial s}F^{\mu\nu}.
\label{eq:Guv}
\end{equation}

To construct a dyonic black hole solution in asymptotically AdS spacetime,
we take the following ansatz for the metric and the gauge pential 
\begin{align}
ds^{2} & =-f\left( r\right) dt^{2}+\frac{1}{f\left( r\right) }%
dr^{2}+r^{2}d\Omega_{d-2}^{2}\text{,}  \label{eq:metric} \\
A & =A_{t}\left( r\right) dt-h\left( {\displaystyle\prod\limits_{i=1}^{d-4}}
\sin^{2}\theta_{i}\right) \cos\theta_{d-3}d\theta_{d-2},  \label{eq:ansatz}
\end{align}
where $d\Omega_{d-2}^{2}$ is the metric of $\left( d-2\right) -$sphere (only
consider the case of positive constant curvature, i.e. $k=1$),%
\begin{align}
d\Omega_{1}^{2} & =d\theta_{1}^{2}, \\
d\Omega_{n+1}^{2} & =d\Omega_{n}^{2}+\left( {\displaystyle%
\prod\limits_{i=1}^{n}} \sin^{2}\theta_{i}\right) d\theta_{n+1}^{2}.
\end{align}
Then the equations of motion read 
\begin{align}
\frac{\left( d-2\right) }{2}rf\left( r\right) ^{\prime} +\frac{\left(
d-2\right) \left( d-3\right) }{2}[f(r)-1] -\frac{\left( d-2\right) \left(
d-1\right) r^{2}}{2l^{2}} & =\frac{r^{2}}{2}\left[ \mathcal{L}\left(
s\right) \mathcal{+}G^{rt}A_{t}^{\prime}\left( r\right) \right] ,
\label{eq:rrEOM} \\
\partial_{r}\left( r^{d-2}G^{rt}\right) & =0\text{,}  \label{eq:NLEDEOM} \\
\partial_{\theta_{d-3}}\left( \sin\theta_{d-3}G^{\theta_{d-3}\theta_{d-2}
}\right) & =0\text{,}  \label{eq:NLEDEOMh}
\end{align}
and plugging eqn. $\left( \ref{eq:ansatz}\right) $ into eqn. $\left( \ref%
{eq:maxwell}\right) $ and eqn. $\left( \ref{eq:Guv}\right) $ results in 
\begin{equation}
s=\frac{A_{t}^{\prime^{2}}\left( r\right) }{2}-\frac{h^{2}}{2r^{4}}\text{
and }G^{rt}=-\frac{\partial\mathcal{L}\left( s\right) }{\partial s}%
A_{t}^{\prime}\left( r\right) \text{.}  \label{eq:sGrt}
\end{equation}
Eqn. $\left( \ref{eq:NLEDEOMh}\right) $ can result in $\partial
_{\theta_{d-3}}h=0$ and the rest equations of motion can be derived from
eqn. $\left( \ref{eq:rrEOM}\right) $ and eqn. $\left( \ref{eq:NLEDEOM}%
\right) .$ Now $A_{t}^{\prime}\left( r\right) $ can be determined by eqn. $%
\left( \ref{eq:NLEDEOM}\right) $, which leads to%
\begin{equation}
\frac{\partial\mathcal{L}\left( s\right) }{\partial s}A_{t}^{\prime}\left(
r\right) =\frac{q}{r^{d-2}},  \label{eq:QAt}
\end{equation}
where $q$ is a constant. Moreover integrating eqn. $\left( \ref{eq:rrEOM}%
\right) $, we have%
\begin{equation}
f\left( r\right) =1+\frac{r^{2}}{l^{2}}-\frac{m}{r^{d-3}}-\frac{1}{\left(
d-2\right) r^{d-3}}\int_{r}^{\infty}drr^{d-2}\left[ \mathcal{L}\left( \frac{%
A_{t}^{\prime^{2}}\left( r\right) }{2}-\frac{h^{2}}{2r^{4}}\right) \mathcal{-%
}\frac{q}{r^{d-2}}A_{t}^{\prime}\left( r\right) \right] ,  \label{eq:fd}
\end{equation}
where $m$ is a constant. At the horizon $r=r_{+}$, the Hawking temperature
of the black hole is given by%
\begin{equation}
T=\frac{f^{\prime}\left( r_{+}\right) }{4\pi},
\end{equation}
so one can have 
\begin{equation}
T=\frac{1}{4\pi r_{+}}\left\{ d-3+\frac{\left( d-1\right) r_{+}^{2}}{l^{2}}+%
\frac{1}{d-2}r_{+}^{2}\left[ \mathcal{L}\left( \frac{A_{t}^{\prime^{2}}%
\left( r\right) }{2}-\frac{h^{2}}{2r^{4}}\right) \mathcal{-}\frac {q}{r^{d-2}%
}A_{t}^{\prime}\left( r_{+}\right) \right] \right\} ,  \label{eq:HT}
\end{equation}
which results from plugging $f\left( r_{+}\right) =0$ into eqn. $\left( \ref%
{eq:rrEOM}\right) .$

Then the electric charge is \cite{Rasheed:1997ns}%
\begin{equation}
Q=\int_{S}\bar{F}=\int\left( {\displaystyle\prod\limits_{1}^{d-2}}
d\theta_{i}\right) \bar{F}=\int\left( {\displaystyle\prod\limits_{i=1}^{d-2}}
d\theta_{i}\right) \sqrt{-g}\frac{q}{r^{d-2}}=\omega_{d-2}q,
\label{eq:anydQ}
\end{equation}
where%
\begin{equation}
\bar{F}=\frac{\partial\mathcal{L}}{\partial s}\left( \ast F\right) ,
\end{equation}
with $\omega_{d-2}$ being the volume of the unit $\left( d-2\right) $-sphere:%
\begin{equation}
\omega_{d-2}=\frac{2\pi^{\frac{d-1}{2}}}{\Gamma\left( \frac{d-1}{2}\right) }.
\end{equation}

Moreover, the mass can be extracted by comparison to a reference background,
e.g., vacuum AdS. So the mass can be determined by the Komar integral%
\begin{equation}
M=\frac{d-2}{8\pi\left( d-3\right) }\int_{\partial\Sigma}dx^{d-2}\sqrt{%
\gamma^{\prime}}\left( \sigma_{\mu}n_{\nu}\nabla^{\mu}K^{\nu}\right) -M_{%
\text{AdS}},
\end{equation}
where $K^{\mu}$ is the Killing vector associated with $t$, and $M_{\text{AdS}%
}$ is Komar integral associated with $K^{\mu}$ for vacuum AdS space%
\begin{equation}
M_{\text{AdS}}=\frac{d-2}{8\pi\left( d-3\right) }\int_{\partial\Sigma
}dx^{d-2}\sqrt{\gamma^{\prime}}\left( \frac{r}{l^{2}}\right) ,
\label{eq:MAdS}
\end{equation}
and $\gamma^{\prime}$ is the induced metric of $\partial\Sigma$, which is
the boundary of $\Sigma$. $\sigma_{\mu}$ is the unit normal vector of $%
\Sigma $ and $n_{\mu}$ is the unit outward-pointing normal vector. Setting $%
\Sigma$ and $\partial\Sigma$ are a constant-$t$ hypersurface and a $\left(
d-2\right) $-sphere at $r=\infty$.

Using%
\begin{align}
\sigma _{\mu }& =(-f^{\frac{1}{2}},0,0,0......),\text{ } \\
n_{\mu }& =(0,f^{-\frac{1}{2}},0,0......)\text{,}
\end{align}%
one can have 
\begin{equation}
\sigma _{\mu }n_{\nu }\nabla ^{\mu }K^{\nu }=\frac{1}{2}f^{\prime }(r).
\end{equation}%
It is shown that in some case, which is 
\begin{equation}
d-1<4p<2d-2\text{,}  \label{eq:Mcond}
\end{equation}%
one can have%
\begin{equation}
\frac{1}{2}f^{\prime }(r)=\frac{r}{l^{2}}+\frac{\left( d-3\right) m}{2r^{d-2}%
}+O(r^{2-d})
\end{equation}%
at spatial infinity, whether it hold or not is determined by the
relationship of power exponent $p$ and dimension $d$. When it hold, we have%
\begin{equation}
M=\frac{d-2}{16\pi }\omega _{d-2}m.  \label{eq:anydmass}
\end{equation}

In the following, we study the thermodynamics of the dyonic PM AdS black
hole solution in the extended phase space, where the cosmological constant
is interpreted as thermodynamic pressure and treated as a thermodynamic
variable in its own right. The mass of the black hole is no longer regarded
as internal energy, it is identified with the chemical enthalpy.

In terms of the horizon radius $r_{+},$ the mass can be rewritten as

\begin{equation}
M=\frac{d-2}{16\pi}\omega_{d-2}\left\{ r_{+}^{d-3}+\frac{r_{+}^{d-1}}{l^{2}}-%
\frac{1}{d-2}\int_{r_{+}}^{\infty}drr^{d-2}\left[ \mathcal{L}\left( \frac{%
A_{t}^{\prime2}(r)}{2}-\frac{h^{2}}{2r^{4}}\right) -A_{t}^{\prime }\left(
r\right) \frac{q}{r^{d-2}}\right] \right\}  \label{eq:anydm}
\end{equation}
where we have used eqn. $\left( \ref{eq:anydmass}\right) $.

Adding that the Gibbs free energy $F$ can be expressed by the Euclidean
action $S^{E}$ \cite{Wang:2018xdz}:%
\begin{equation}
F=M-TS,
\end{equation}
where the entropy of the black hole is one-quarter of the horizon area%
\begin{equation}
S=\frac{r_{+}^{d-2}\omega_{d-2}}{4}.  \label{eq:entropy}
\end{equation}
We have expressed thermodynamics quantities $\left( F,M\text{ and }S\right) $
as the functions of the horizon radius $r_{+}$, $q$ (proportional to the
electric charge $Q$), $h$ (associated with magnetic charge) and the AdS
radius $l$ $\left( \text{the pressure }P=\left( d-1\right) \left( d-2\right)
/l^{2}\right) $. Now we need to express the thermodynamics quantities in
terms of $T$, $q$, $h$ and $P$ by solving the equation of state for%
\begin{equation}
r_{+}=r_{+}\left( T,q,l,h\right) .
\end{equation}
So we first rescale the $T$, which becomes%
\begin{equation}
\tilde{T}=\frac{1}{4\pi\tilde{r}_{+}}\left\{ d-3+\left( d-1\right) \tilde{r}%
_{+}^{2}+\frac{\tilde{r}_{+}^{2}}{d-2}\left[ \left( \frac{\tilde {A}%
_{t}^{\prime2}(r_{+})}{2}-\frac{\tilde{h}^{2}}{2\tilde{r}_{+}^{4}}\right)
^{p}-\tilde{A}_{t}^{\prime}\left( r_{+}\right) \frac{\tilde{q}}{\tilde {r}%
_{+}^{d-2}}\right] \right\} ,  \label{eq:Ttildal}
\end{equation}
where%
\begin{equation}
\tilde{r}_{+}=r_{+}l^{-1},\text{ }\tilde{q}=ql^{-\frac{1}{p}-d+4},\text{ }%
\tilde{A}_{t}^{^{\prime}}(r_{+})=l^{\frac{1}{p}}A_{t}^{\prime}(r_{+}),\text{ 
}\tilde{h}=hl^{\frac{1}{p}-2},\text{ }\tilde{T}=Tl,
\end{equation}
and $p$ is the power of%
\begin{equation}
\mathcal{L}\left( s\right) =s^{p}.
\end{equation}
Then $\tilde{A}_{t}^{\prime}\left( r_{+}\right) $ is determined by%
\begin{equation}
\left( p\frac{\tilde{A}_{t}^{\prime2}(r_{+})}{2}-\frac{\tilde{h}^{2}}{2%
\tilde{r}_{+}^{4}}\right) ^{p-1}\tilde{A}_{t}^{\prime}\left( r_{+}\right) =%
\frac{\tilde{q}}{\tilde{r}_{+}^{d-2}},  \label{eq:QAtTtildal}
\end{equation}
which usually cannot be solved analytically and has multiple solutions when $%
p$ is large. After that, solving eqn. $\left( \ref{eq:Ttildal}\right) $, $%
\tilde{r}_{+}$ can be expressed as a function of $\tilde{T}$, $\tilde{q}$
and $\tilde{h}$: $\tilde{r}_{+}=\tilde{r}_{+}(\tilde{T},\tilde{q},\tilde{h})$%
. With $\tilde{r}_{+}=\tilde{r}_{+}(\tilde{T},\tilde{q},\tilde{h}_{i})$, one
can express the thermodynamic quantities in terms of $\tilde{T},\tilde{q}$
and $\tilde{h}$, e.g., the Gibbs free energy is given by%
\begin{equation}
\tilde{F}\equiv F/l^{d-3}=\tilde{F}(\tilde{T},\tilde{q},\tilde{h}).
\end{equation}

\begin{figure}[ptb]
\begin{center}
\subfigure[{~\scriptsize Branches around a local minimum of $\tilde{T}=\tilde{T}_{\min}$.}]{
\includegraphics[width=0.95\textwidth]{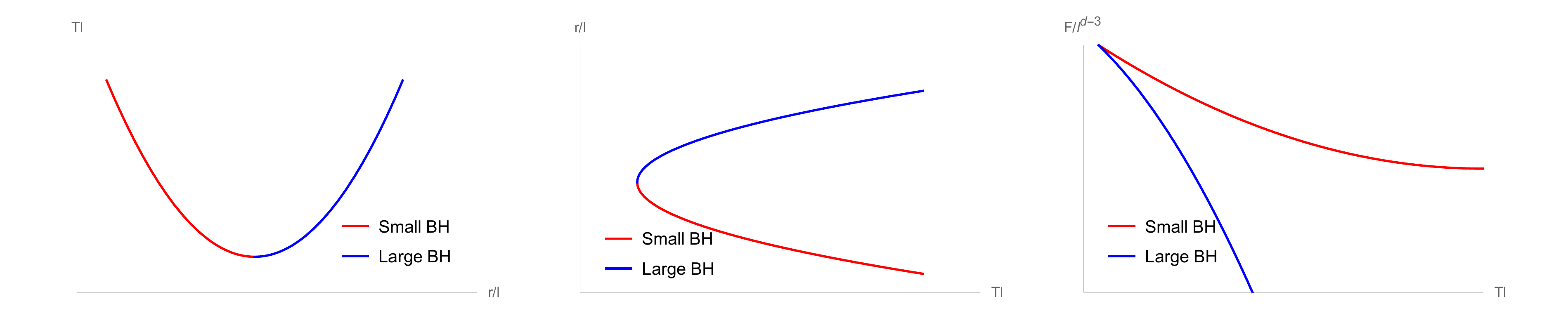}\label{fig:M:a}} 
\subfigure[{~\scriptsize Branches around a local maximum of $\tilde{T}=\tilde{T}_{\max}$ when $\tilde{T}=\tilde{T}(\tilde{r}_{+};\tilde
	{q},\tilde{h})$ is multivalued.}]{
\includegraphics[width=0.95\textwidth]{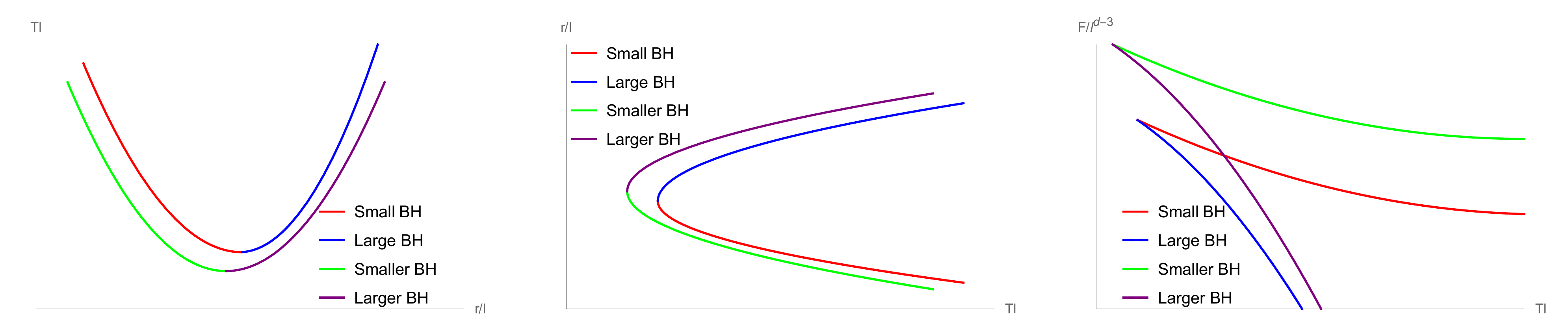}\label{fig:M:b}}
\end{center}
\caption{{\protect\footnotesize Branches of black holes around local
extremums of $\tilde{T}=\tilde{T}_{\min}$. Right panels: Gibbs free energy
vs temperature. The blue branches are thermodynamically preferred and
thermally stable. The red ones are thermally unstable. Below column: the
case of} {\protect\footnotesize more than one }${\protect\footnotesize 
\tilde{T}(\tilde{r}}_{+}{\protect\footnotesize ,\tilde{q},\tilde{h})}$ 
{\protect\footnotesize with some fixed }${\protect\footnotesize \tilde{q}}$ 
{\protect\footnotesize and }${\protect\footnotesize \tilde{h}.}$}
\label{fig:M}
\end{figure}

The rich phase structure of the black hole comes from solving eqn. $\left( %
\ref{eq:Ttildal}\right) $, i.e., $\tilde{T}=\tilde{T}(\tilde{r}_{+},\tilde {q%
},\tilde{h})$, for $\tilde{r}_{+}$. If $\tilde{T}(\tilde{r}_{+},\tilde {q},%
\tilde{h})$ is a monotonic function with respect to $\tilde{r}_{+}$ for some
values of $\tilde{q}$ and $\tilde{h}$, there would be only one branch for
the black hole. More often, with fixed $\tilde{q}$ and $\tilde{h}$, there
exists a local minimum/maximum for $\tilde{T}(\tilde{r}_{+},\tilde{q},\tilde{%
h})$ at $\tilde{r}_{+}=\tilde{r}_{+,\min}/\tilde{r}_{+}=\tilde {r}_{+,\max}$%
. In this case, there are more than one branch for the black hole. In FIG. %
\ref{fig:M:a}, we plot two branches, namely small BH and large BH, around a
local minimum of $\tilde{T}=\tilde{T}_{\min}$. The Gibbs free energy of
these two branches is displayed in the right panel of FIG. \ref{fig:M:a}.
Since $\partial\tilde{F}(\tilde{T},\tilde{q},\tilde{h})/\partial\tilde {T}%
=-4\pi^{6}\tilde{r}_{+}^{6}$, the upper branch is small BH while the lower
one is large BH, which means that the large BH branch is thermodynamically
preferred. Similarly, there are also two branches around a local maximum of $%
\tilde{T}$. In this case the upper/lower branch is large/small BH since it
has more/less negative slope and the small BH branch is thermodynamically
preferred in this case. In general, one might need to figure out how the
existence of local extremums depends on values of $\tilde{Q}$ and $\tilde{h}$
to study the phase structure of the black hole.

Moreover, since there are more than one solution by solving the eqn. $\left( %
\ref{eq:QAtTtildal}\right) $ for some values of $\tilde{q}$ and $\tilde{h}$,
there are more than one $\tilde{T}(\tilde{r}_{+},\tilde{q},\tilde{h})$ with
some fixed $\tilde{q}$ and $\tilde{h}$, which means more one set of $\tilde {%
r}_{+i}=\tilde{r}_{+i}(\tilde{T},\tilde{q},\tilde{h})$. And every set of $%
\tilde{r}_{+i}(\tilde{T},\tilde{q},\tilde{h})$ maybe still have many
branches. In FIG. \ref{fig:M:b}, we show a possible simple case, every panel
is corresponding to those of FIG. \ref{fig:M:a}. And of couse it could be
more complicated.

After the black hole's branches are obtained, it is easy to check their
thermodynamic stabilities against thermal fluctuations. The thermal
stability of the branch follows from the specific heat $C>0$. The specific
heat we need is%
\begin{equation}
C_{Q,h,P}=T\left( \frac{\partial S}{\partial T}\right)
_{Q,h,P}=l^{d-2}\left( d-2\right) \tilde{T}\frac{\tilde{r}%
_{+}^{d-3}\omega_{d-2}}{4}\left( \frac{\partial\tilde{r}_{+}}{\partial\tilde{%
T}}\right) ,
\end{equation}
since $\omega_{d-2}$ is also positive, the sign of $\tilde{T}^{\prime}\left( 
\tilde{r}_{+}\right) $ determines the thermodynamic stabilities.

\section{8-d Dyonic PM AdS Black Hole with p=2}

\label{Sec:PMI2}

In this section, we focus on a specific example. For simplicity, we consider
the case that the power exponent is $2$, and we consider $8-d$ spacetime in
order to satisfying the condition of eqn. $\left( \ref{eq:Mcond}\right) $.

\bigskip When $d=8$, $p=2$ the mass becomes%
\begin{equation}
M=\frac{2}{5}\pi^{2}\left\{ r_{+}^{5}+\frac{r_{+}^{7}}{l^{2}}-\frac{1}{6}%
\int_{r_{+}}^{\infty}drr^{6}\left[ \mathcal{L}\left( \frac{A_{t}^{\prime
2}(r)}{2}-\frac{h^{2}}{2r^{4}}\right) -A_{t}^{\prime}\left( r\right) \frac{q%
}{r^{6}}\right] \right\}  \label{eq:Mass}
\end{equation}
and the electric charge%
\begin{equation}
Q=\frac{16\pi^{3}}{15}q,  \label{eq:Q}
\end{equation}
where we have used eqn. $\left( \ref{eq:anydm}\right) $ and eqn. $\left( \ref%
{eq:anydQ}\right) .$ And the eqn. $\left( \ref{eq:Ttildal}\right) $ and eqn. 
$\left( \ref{eq:entropy}\right) $ become 
\begin{equation}
\tilde{T}=\frac{1}{4\pi\tilde{r}_{+}}\left\{ 5+7\tilde{r}_{+}^{2}+\frac{%
\tilde{r}_{+}^{2}}{6}\left[ \mathcal{L}\left( \frac{\tilde{A}%
_{t}^{\prime2}(r_{+})}{2}-\frac{\tilde{h}^{2}}{2\tilde{r}_{+}^{4}}\right) -%
\tilde{A}_{t}^{\prime}\left( r_{+}\right) \frac{\tilde{q}}{\tilde{r}_{+}^{6}}%
\right] \right\} ,
\end{equation}%
\begin{equation}
S=\frac{4}{15}\pi^{3}r_{+}^{6},
\end{equation}
where 
\begin{equation}
\tilde{r}_{+}=r_{+}l^{-1},\text{ }\tilde{q}=ql^{-4.5},\text{ }\tilde{A}%
_{t}^{^{\prime}}(r_{+})=l^{0.5}A_{t}^{\prime}(r_{+}),\text{ }\tilde {h}%
=hl^{-1.5},\text{ }\tilde{T}=Tl.
\end{equation}
Moreover, the Gibbs free energy is given by%
\begin{equation}
F=M-TS
\end{equation}
and%
\begin{equation}
\tilde{F}\equiv F/l^{5}=\tilde{F}(\tilde{T},\tilde{q},\tilde{h}).
\end{equation}

The case of $p=2$ is described by the Lagrangian density%
\begin{equation}
\mathcal{L}\left( s\right) =s^{2}\text{,}  \label{eq:PL}
\end{equation}
and solving eqn. $\left( \ref{eq:QAtTtildal}\right) $ for $\tilde{A}%
_{t}^{\prime}\left( r\right) $ gives 
\begin{equation}
\tilde{A}_{ti}^{\prime}\left( r_{+}\right) =\frac{2\tilde{h}}{\sqrt{3}\tilde{%
r}_{+}^{2}}\cos\left[ \frac{1}{3}\arccos\left( \frac{3\sqrt{3}\tilde{q}}{2%
\tilde{h}^{3}}\right) -\frac{2\pi}{3}\left( i-1\right) \right] \text{, }%
i=1,2,3\text{,}
\end{equation}
where $\tilde{A}_{t2}^{\prime}\left( r_{+}\right) $ and $\tilde{A}%
_{t3}^{\prime}\left( r_{+}\right) $ exist only if ${3\sqrt{3}\tilde{q}}/{2%
\tilde{h}^{3}}\leq1$ for $\tilde{h}\tilde{q}>0$; $\tilde{A}_{t2}^{\prime
}\left( r_{+}\right) $ and $\tilde{A}_{t1}^{\prime}\left( r_{+}\right) $
exist only if ${3\sqrt{3}\tilde{q}}/{2\tilde{h}^{3}}\geq-1$ for $\tilde {h}%
\tilde{q}<0.$

The equations of state $\left( \ref{eq:Ttildal}\right) $ become 
\begin{align}
\tilde{T}_{i} & =\frac{1}{4\pi\tilde{r}_{+}}\left\{ 5+7\tilde{r}_{+}^{2}+%
\frac{\tilde{r}_{+}^{2}}{6}\left[ \left( \frac{\tilde{A}_{ti}^{\prime
2}(r_{+})}{2}-\frac{\tilde{h}^{2}}{2\tilde{r}_{+}^{4}}\right)
^{2}-A_{ti}^{\prime}\left( r_{+}\right) \frac{\tilde{q}}{\tilde{r}_{+}^{6}}%
\right] \right\} \\
& =\frac{1}{4\pi\tilde{r}_{+}}\left\{ 5+7\tilde{r}_{+}^{2}+\frac{\tilde {r}%
_{+}^{2}}{6}\left[ \left( \frac{2\tilde{h}^{2}}{3\tilde{r}_{+}^{4}}%
C_{i}^{2}\left( \tilde{h},\tilde{q}\right) -\frac{\tilde{h}^{2}}{2\tilde {r}%
_{+}^{4}}\right) ^{2}-\frac{2\tilde{h}\tilde{q}}{\sqrt{3}r_{+}^{8}}%
C_{i}\left( \tilde{h},\tilde{q}\right) \right] \right\} \\
& =\frac{1}{4\pi\tilde{r}_{+}}\left[ 5+7\tilde{r}_{+}^{2}+\frac{\tilde {h}%
^{4}}{6\tilde{r}_{+}^{6}}D_{i}\left( \tilde{h},\tilde{q}\right) \right] ,
\label{eq:PLT}
\end{align}
and we have defined two functions that are independent of $\tilde{r}_{+}$
for later use 
\begin{align}
C_{i}\left( \tilde{h},\tilde{q}\right) & =\cos\left[ \frac{1}{3}%
\arccos\left( \frac{3\sqrt{3}\tilde{q}}{2\tilde{h}^{3}}\right) -\frac{2\pi }{%
3}\left( i-1\right) \right] \text{(}i=1,2,3\text{),} \\
D_{i}\left( \tilde{h},\tilde{q}\right) & =\left[ \left( \frac{%
4C_{i}^{2}\left( \tilde{h},\tilde{q}\right) }{3}-1\right) \frac{1}{2}\right]
^{2}-C_{i}\left( \tilde{h},\tilde{q}\right) \frac{2\tilde{q}}{\sqrt{3}\tilde{%
h}^{3}}.
\end{align}
Setting ${3\sqrt{3}\tilde{q}}/{2\tilde{h}^{3}}\equiv x$, we have 
\begin{align}
C_{i}\left( x\right) & \equiv\cos\left[ \frac{1}{3}\arccos x-\frac{2\pi }{3}%
\left( i-1\right) \right] \text{(}i=1,2,3\text{)}, \\
D_{i}\left( x\right) & \equiv\left\{ \left[ \left( \frac{4C_{i}^{2}\left(
x\right) }{3}-1\right) \frac{1}{2}\right] ^{2}-xC_{i}\left( x\right) \frac{4%
}{9}\right\} ,
\end{align}
and we can see that $\tilde{h}^{4}D_{i}\left( x\right) $ completely
determines the dependence of $\tilde{T}_{i}\left( \tilde{r}_{+}\right) $ on $%
\tilde{r}$,%
\begin{equation}
\tilde{T}_{i}\left( \tilde{r}_{+};\tilde{h},x\right) =\frac{1}{4\pi\tilde {r}%
_{+}}\left[ 5+7\tilde{r}_{+}^{2}+\frac{\tilde{h}^{4}}{6\tilde{r}_{+}^{6}}%
D_{i}\left( x\right) \right] ,
\end{equation}
which also tells us that three curves of $\tilde{T}_{i}\left( \tilde{r}_{+};%
\tilde{h},x\right) $ are never intersect with fixed $\tilde{h},x$. Moreover,
we can write the rescaled Gibbs free energy as%
\begin{equation}
\tilde{F}=\frac{12\pi^{2}\tilde{r}_{+}^{5}}{5}+\frac{16\pi^{2}\tilde{r}%
_{+}^{7}}{5}-\frac{28}{15}\pi^{3}\tilde{r}_{+}^{6}\tilde{T},
\end{equation}
and it is also easy to see that three curves of $F\left( \tilde{T}_{i};%
\tilde{h},x\right) $ are never intersect with fixed $\tilde{h},x$.\newline
\begin{figure}[ptb]
\begin{center}
\subfigure[{~ \scriptsize $C\left(  x\right)$}]{
\includegraphics[width=0.48\textwidth]{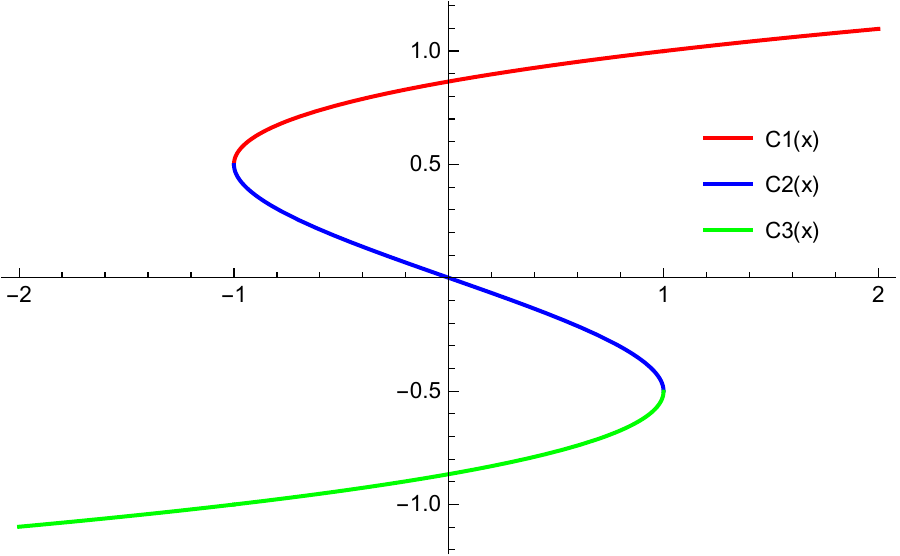}\label{fig:CD:a}} 
\subfigure[{~ \scriptsize  $D\left(  x\right)$}]{
\includegraphics[width=0.48\textwidth]{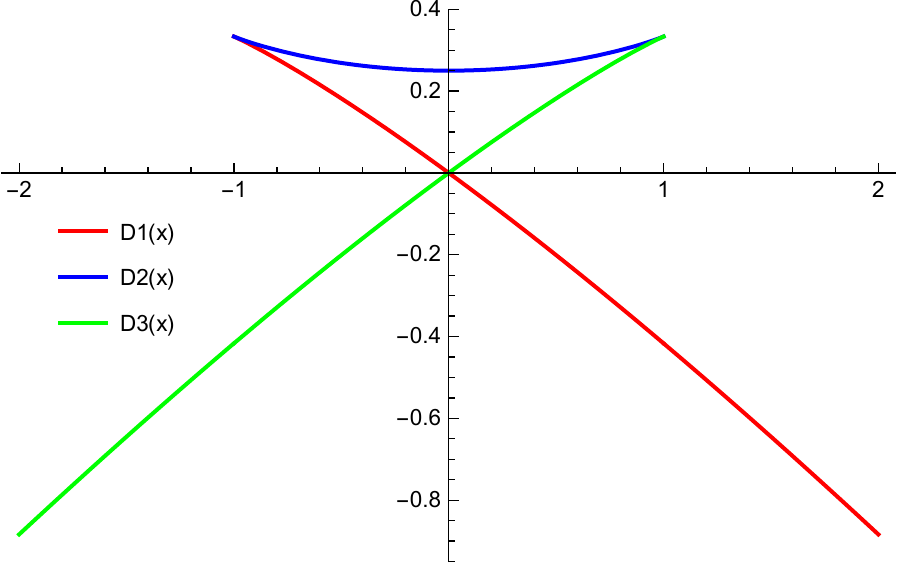}\label{fig:CD:b}}
\end{center}
\caption{${\protect\footnotesize C(x)}$ {\protect\footnotesize and} $%
{\protect\footnotesize D(x)}$, {\protect\footnotesize where }$x\equiv \frac{3%
\protect\sqrt{3}\tilde{q}}{2\tilde{h}^{3}}.$}
\label{fig:CD}
\end{figure}
To study the behavior of local extremums of $\tilde{T}_{i}\left( \tilde {r}%
_{+}\right) $, we consider the equation $\tilde{T}_{i}^{\prime\prime }\left( 
\tilde{r}_{+}\right) =0$, which becomes%
\begin{equation}
\frac{d^{2}\tilde{T}_{i}}{d\tilde{r}_{+}^{2}}=\frac{10}{4\pi\tilde{r}_{+}^{3}%
}+\frac{7\tilde{h}^{4}}{3\pi\tilde{r}_{+}^{9}}D_{i}\left( \tilde{h},\tilde {q%
}\right) =0.
\end{equation}
For $\tilde{h}\tilde{q}>0$, since $D_{2}\left( \tilde{h},\tilde{q}\right) >0$
and $D_{3}\left( \tilde{h},\tilde{q}\right) >0,$ which is shown in FIG. \ref%
{fig:CD:b}, only $\tilde{T}_{1}^{\prime\prime}\left( \tilde{r}_{+}\right) =0$
has a root and only has one root. And when $\tilde{h}\tilde{q}<0$, only $%
\tilde{T}_{3}^{\prime\prime}\left( \tilde{r}_{+}\right) =0$ has and only has
a root. From FIG. \ref{fig:CD:b}, we can find $D_{1}\left( x\right) $ is
symmetry with $D_{3}\left( x\right) $, which means $\tilde{T}_{1}\left( 
\tilde{r}_{+}\right) /\tilde{T}_{3}\left( \tilde{r}_{+}\right) $ for $\tilde{%
h}\tilde{q}>0$ is identical to $\tilde {T}_{3}\left( \tilde{r}_{+}\right) /%
\tilde{T}_{1}\left( \tilde{r}_{+}\right) $ for $\tilde{h}\tilde{q}<0$. So we
only discuss the case of $\tilde{q}>0$, $\tilde{h}>0$ in the following. With
solutions of $\tilde {T}_{i}^{\prime\prime}\left( \tilde{r}_{+}\right) =0$,
it is easy to analyze the existence of the local extremums of $\tilde{T}%
^{\prime}\left( \tilde {r}_{+}\right) $, results of which are summarized in
Table \ref{tab:1}.

\begin{table}[tbh]
\centering%
\begin{tabular}{|c|c|c|c|c|}
\hline
$\tilde{h}>0,\tilde{q}>0$ & $\tilde{T}^{\prime}\left( 0\right) $ & $\tilde{T}%
^{\prime}\left( +\infty\right) $ & Solution of $\tilde{T}^{\prime\prime}%
\left( \tilde{r}_{+}\right) =0$ & Extremums of $\tilde {T}^{\prime}\left( 
\tilde{r}_{+}\right) $ \\ \hline
$T_{1}$, $\tilde{h}^{4}D_{1}\left( \tilde{h},\tilde{q}\right) >-2\left( 
\frac{15}{28}\right) ^{4}$ & $\infty$ & $\frac{7}{4\pi}$ & $\tilde{r}_{1}>0$
& $\tilde{T}^{\prime}_{min}\left( \tilde{r}_{1}\right) <0$ \\ \hline
$T_{1}$, $\tilde{h}^{4}D_{1}\left( \tilde{h},\tilde{q}\right) <-2\left( 
\frac{15}{28}\right) ^{4}$ & $\infty$ & $\frac{7}{4\pi}$ & $\tilde{r}_{1}>0$
& $\tilde{T}_{min}^{\prime}\left( \tilde{r}_{1}\right) >0$ \\ \hline
$T_{2}$, $x<1$ & $-$$\infty$ & $\frac{7}{4\pi}$ & None, $\tilde{T}%
^{\prime\prime}\left( \tilde{r}_{+}\right) >0$ & None \\ \hline
$T_{3}$, $x<1$ & $-$$\infty$ & $\frac{7}{4\pi}$ & None, $\tilde{T}%
^{\prime\prime}\left( \tilde{r}_{+}\right) >0$ & None \\ \hline
\end{tabular}%
\caption{{\protect\small Solution of $\tilde{T}_{i}^{\prime\prime}\left( 
\tilde{r}_{+}\right) =0$ and the local extremums of $\tilde{T}_{i}^{\prime
}\left( \tilde{r}_{+}\right) $ in various cases.}}
\label{tab:1}
\end{table}

When solving eqn. $\left( \ref{eq:PLT}\right) $ for $\tilde{r}_{+}$ in terms
of $\tilde{T}$, the solution $\tilde{r}_{+}(\tilde{T})$ is often a
multivalued function like what is shown in FIG. \ref{fig:M:b}. The
parameters $\tilde{h}$ and $\tilde{q}$ determine the number of the branches
of $\tilde{r}_{+}(\tilde{T})$ and the phase structure of the black hole. In
what follows, we find four regions in the $\tilde{h}$-$\tilde{q}$ plane if
we only consider the number of branches of $\tilde{r}_{+}(\tilde{T})$,
taking into account for phase transition, Region I has five subregions. Each
region has the distinct behavior of the branches and the phase structure:

\begin{figure}[ptb]
\begin{center}
\subfigure[{~\scriptsize Region I: $h/l^{1.5}=1$ and $q/l^{4.5}=0.02$.}]{
\includegraphics[width=0.48\textwidth]{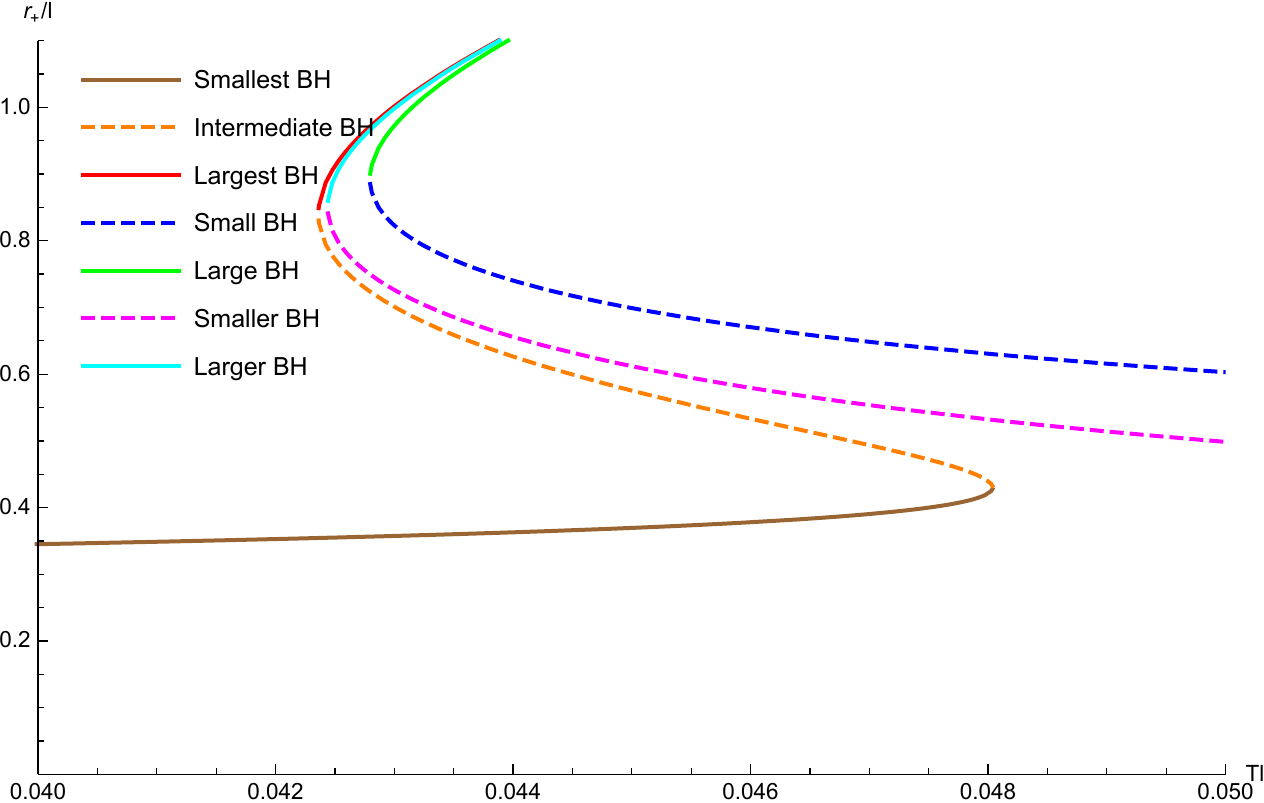}\label{fig:R1:a}} 
\subfigure[{~\scriptsize Region I5:  $h/l^{1.5}=0.5$ and $q/l^{4.5}=0.02$. There is a first order phase transition between Smallest BH and Large BH.}]{
\includegraphics[width=0.48\textwidth]{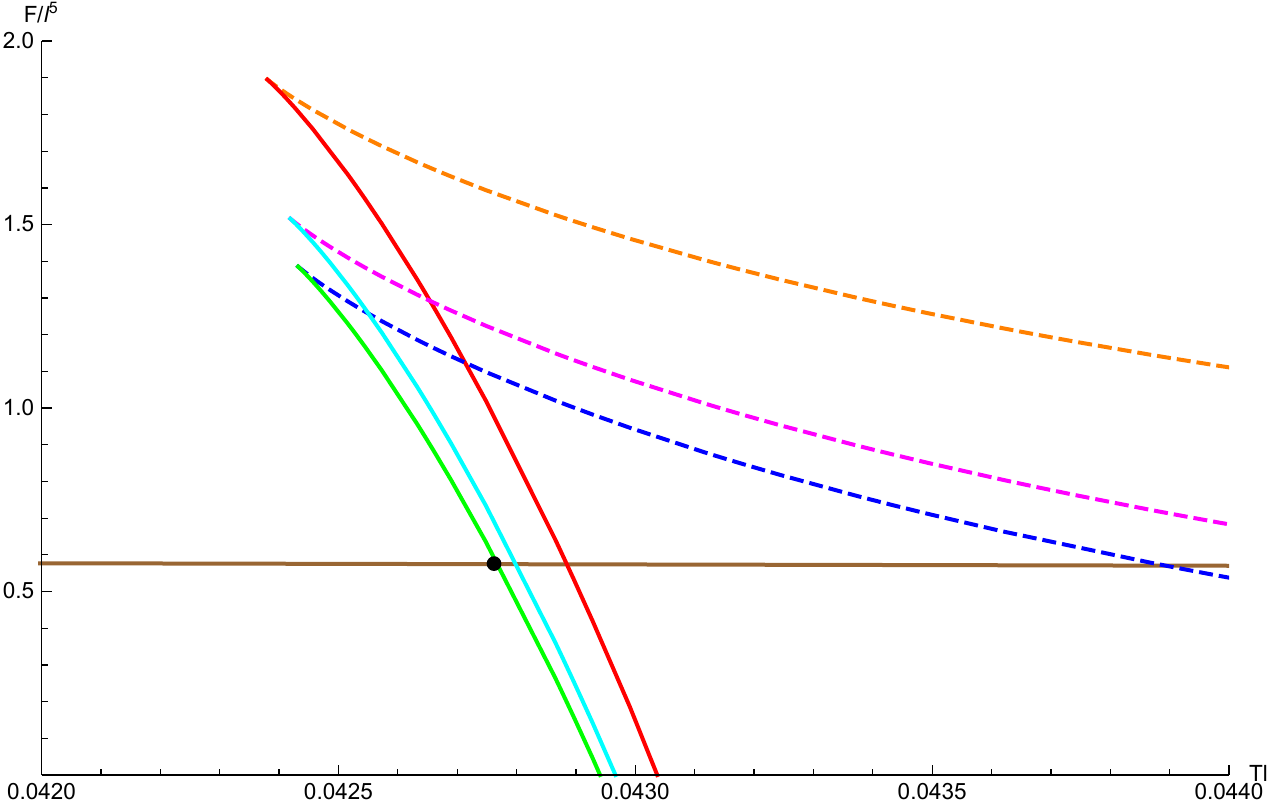}\label{fig:R1:b}} 
\subfigure[{~\scriptsize Region I4:  $h/l^{1.5}=1$ and $q/l^{4.5}=0.001$. There is a zeroth order phase transition between Smallest BH and Large BH.}]{
\includegraphics[width=0.48\textwidth]{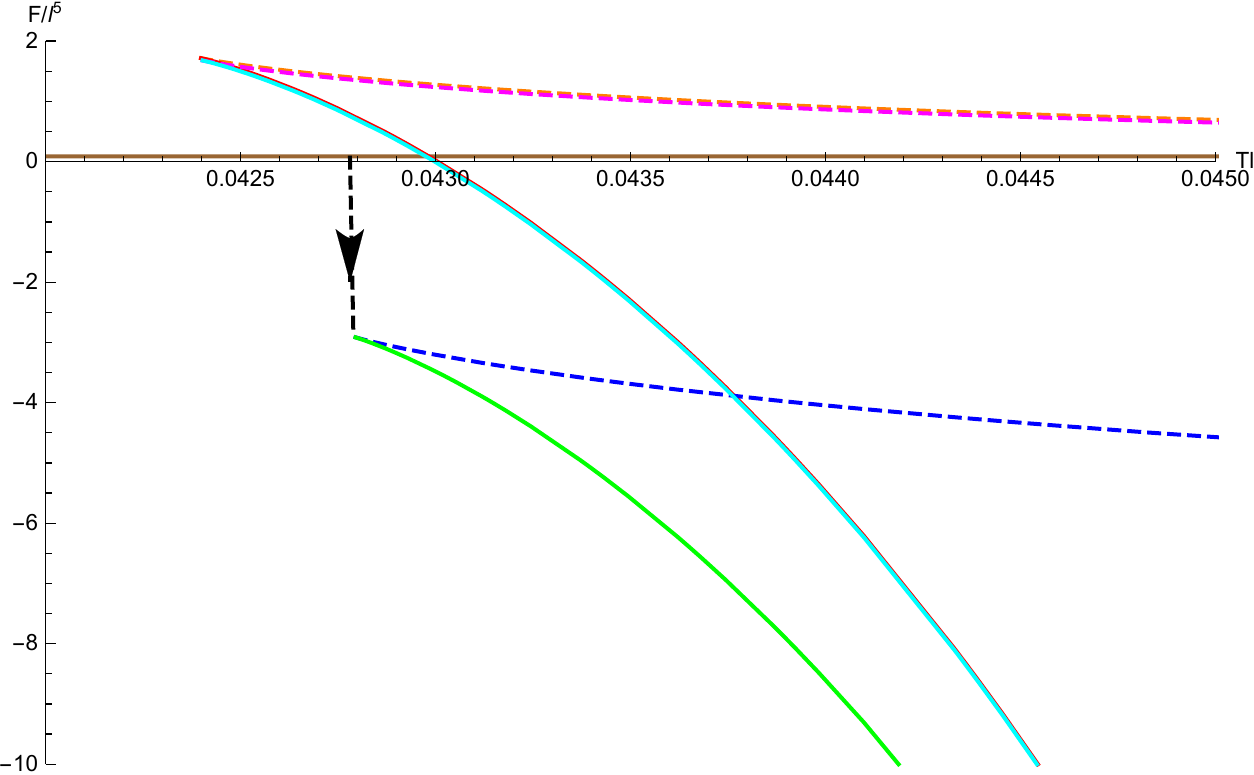}\label{fig:R1:c}} 
\subfigure[{~\scriptsize Region I3:  $h/l^{1.5}=1$ and $q/l^{4.5}=0.02$. There is a first order phase transition between Smallest BH and Larger BH and a zeroth order phase transition between Larger BH and Large BH.}]{
\includegraphics[width=0.48\textwidth]{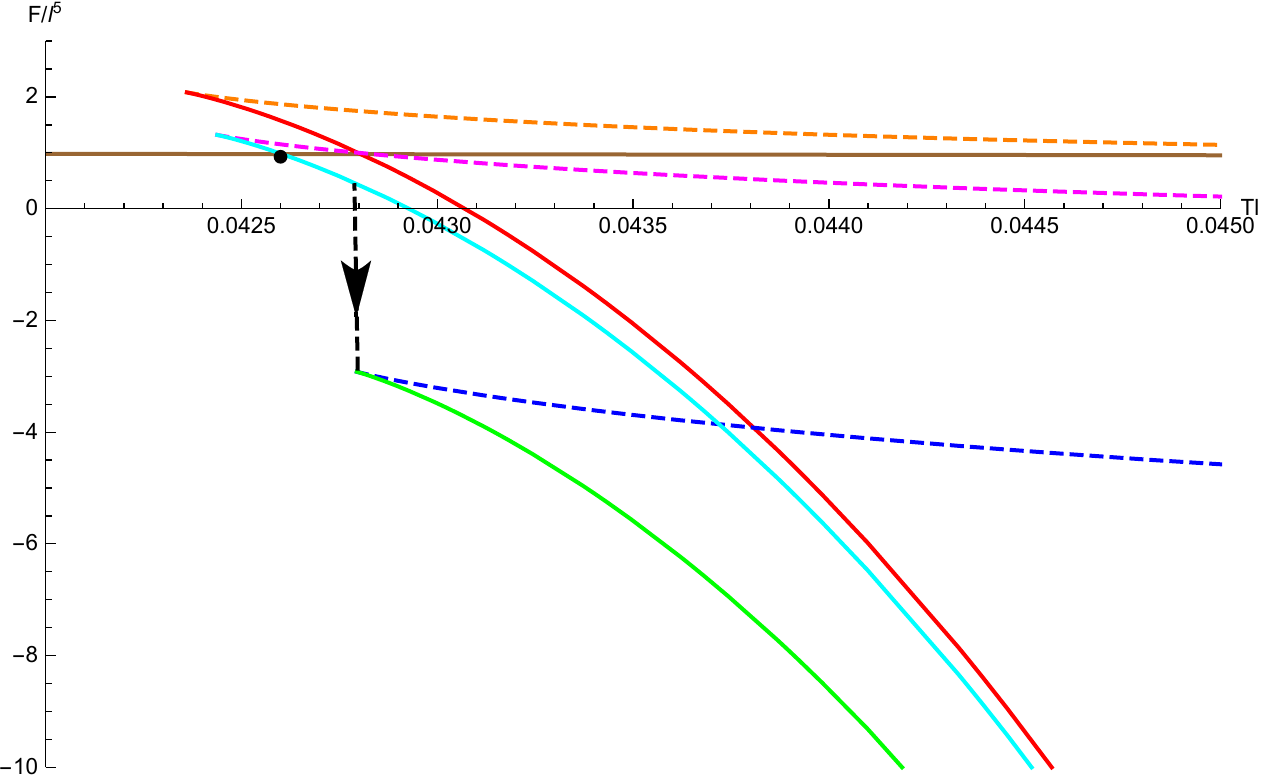}\label{fig:R1:d}} 
\subfigure[{~\scriptsize Region I2:  $h/l^{1.5}=1$ and $q/l^{4.5}=0.05$. There is a zeroth order phase transition between Smallest BH and Larger BH and a zeroth order phase transition between Larger BH and Large BH.}]{
\includegraphics[width=0.48\textwidth]{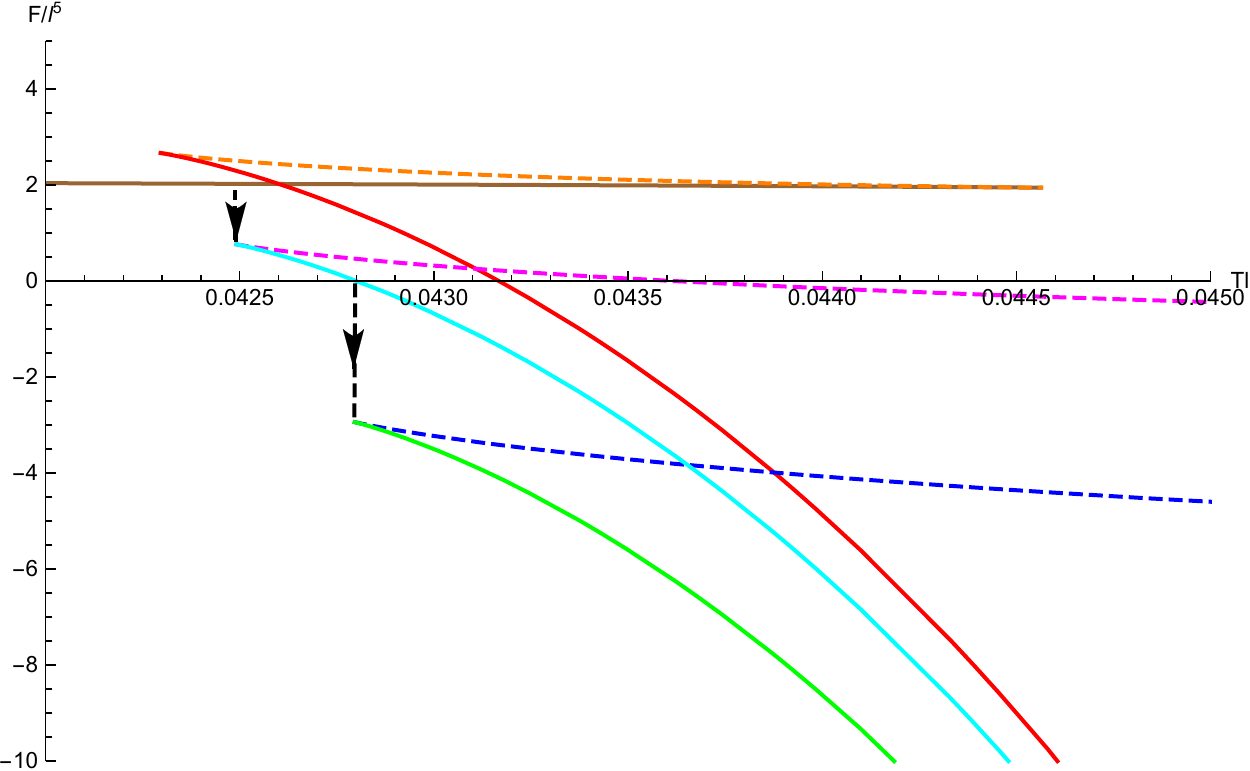}\label{fig:R1:e}} 
\subfigure[{~\scriptsize Region I1: $h/l^{1.5}=1$ and $q/l^{4.5}=0.1$. There is a first order phase transition between Smallest BH and Largest BH, a zeroth order phase transition between Largest BH and Larger BH and a zeroth order phase transition between Larger BH and Large BH.}]{
\includegraphics[width=0.48\textwidth]{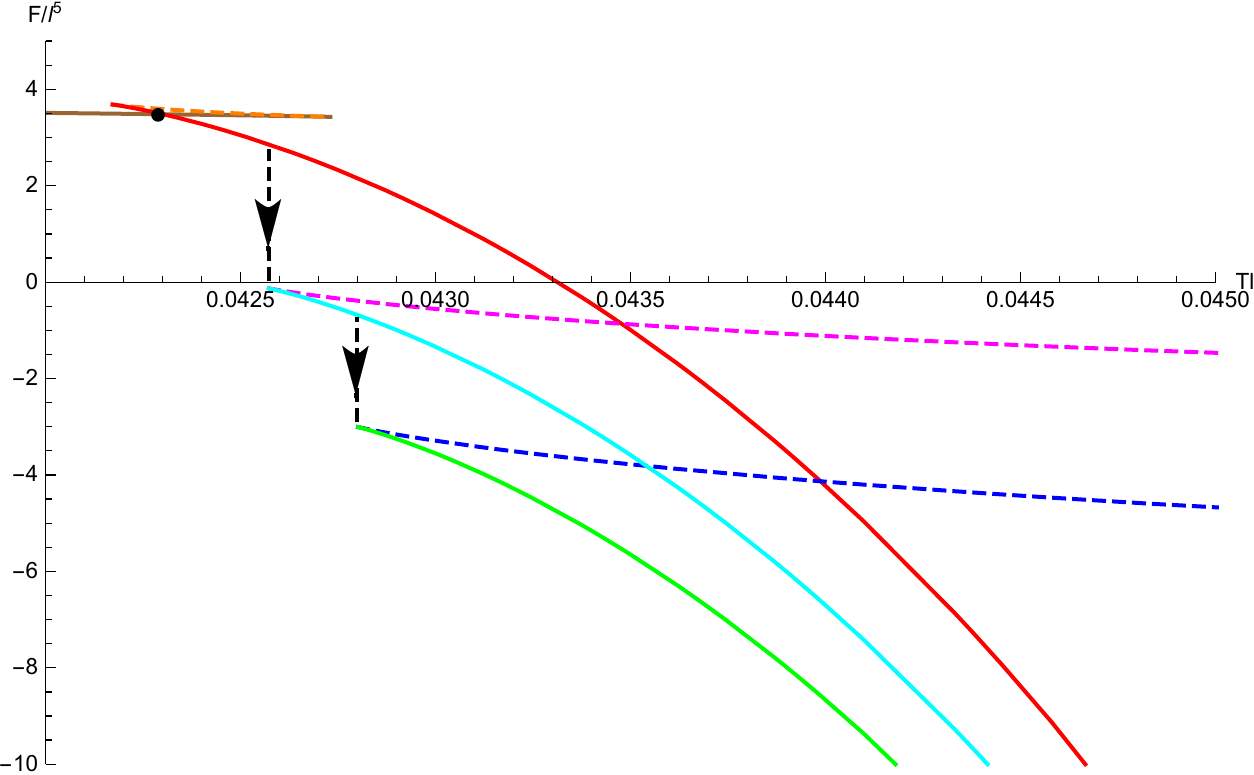}\label{fig:R1:f}}
\end{center}
\caption{{\protect\footnotesize Plot of $\tilde{r}_{+}$, $\tilde{F}$ against 
$\tilde{T}$ for PL-AdS black holes in Region I. the first of this is $\tilde{%
r}_{+}-\tilde{T}$ of Regions I. The shape of $\tilde{r}_{+}-\tilde{T}$ is
same in these five subregions.}}
\label{fig:R1}
\end{figure}

\begin{itemize}
\item Region I: $x<1$ and $\tilde{T}_{1}^{\prime}\left( \tilde{r}_{1}\right)
<0$, where $\tilde{r}_{1}$ is the solution of $\tilde{T}_{1}^{\prime\prime
}\left( \tilde{r}_{+}\right) =0$. In this region, $\tilde{T}_{1}^{\prime
}\left( \tilde{r}_{+}\right) =0$ has two solutions $\tilde{r}_{+}=$\ $\tilde{%
r}_{\text{max}}$ and\ $\tilde{r}_{+}=\tilde{r}_{\text{min}}$ with $\tilde{r}%
_{\text{max}}<\tilde{r}_{1}<\tilde{r}_{\text{min}}$. Since $\tilde{T}%
_{1}\left( +\infty\right) =+\infty$, $\tilde{T}_{1}\left( \tilde{r}%
_{+}\right) $ has a local maximum of $\tilde{T}_{1\text{max}}=\tilde{T}%
_{1}\left( \tilde{r}_{\text{max}}\right) $ at $\tilde{r}_{+}=$\ $\tilde{r}_{%
\text{max}}$ and a local minimum of $\tilde{T}_{1\text{min}}=\tilde{T}%
_{1}\left( \tilde{r}_{\text{min}}\right) $ at $\tilde{r}_{+}=$\ $\tilde{r}_{%
\text{min}}$. There are three branches for $\tilde{r}_{+1}(\tilde{T})$:
smallest BH for $0\leq\tilde{T}_{1}\leq\tilde {T}_{1\text{max}}$,
intermediate BH for $\tilde{T}_{1\text{min}}\leq\tilde {T}_{1}\leq\tilde{T}%
_{1\text{max}}$ and largest BH for $\tilde{T}_{1}\geq$ $\tilde{T}_{1\text{min%
}}$, there are two branches for $\tilde{r}_{+2}(\tilde{T})$: small BH for $%
0\leq\tilde{T}_{2}\leq\tilde{T}_{2\text{min}}$ and large BH for $\tilde{T}%
_{2}\geq$ $\tilde{T}_{2\text{min}}$, and there are also two branches for $%
\tilde{r}_{+3}(\tilde{T})$: smaller BH for $0\leq\tilde {T}_{3}\leq\tilde{T}%
_{3\text{min}}$ and larger BH for $\tilde{T}_{3}\geq$ $\tilde{T}_{3\text{min}%
}$. The dependence of $\tilde{r}_{+}$ to $T$ is displayed in the left panel
of FIG. \ref{fig:R1:a}. The Gibbs free energy of the three branches is
plotted in the follow five panels for different subregions. The smallest BH,
large BH, larger BH and largest BH branches are thermally stable. (When $%
x=1, $ there are three subregions, and they are similar with the case I5,
I2, I1 respectivly, except $\tilde{T}_{2}$ and $\tilde{T}_{2}$ merge into
one.)
\end{itemize}

\begin{figure}[ptb]
\begin{center}
\subfigure[{~\scriptsize Region II:$h/l^{1.5}=1$ and $q/l^{4.5}=0.2$. There is a zeroth order phase transition between Largest BH and Larger BH and a zeroth order phase transition between Larger BH and Large BH.}]{
\includegraphics[width=1\textwidth]{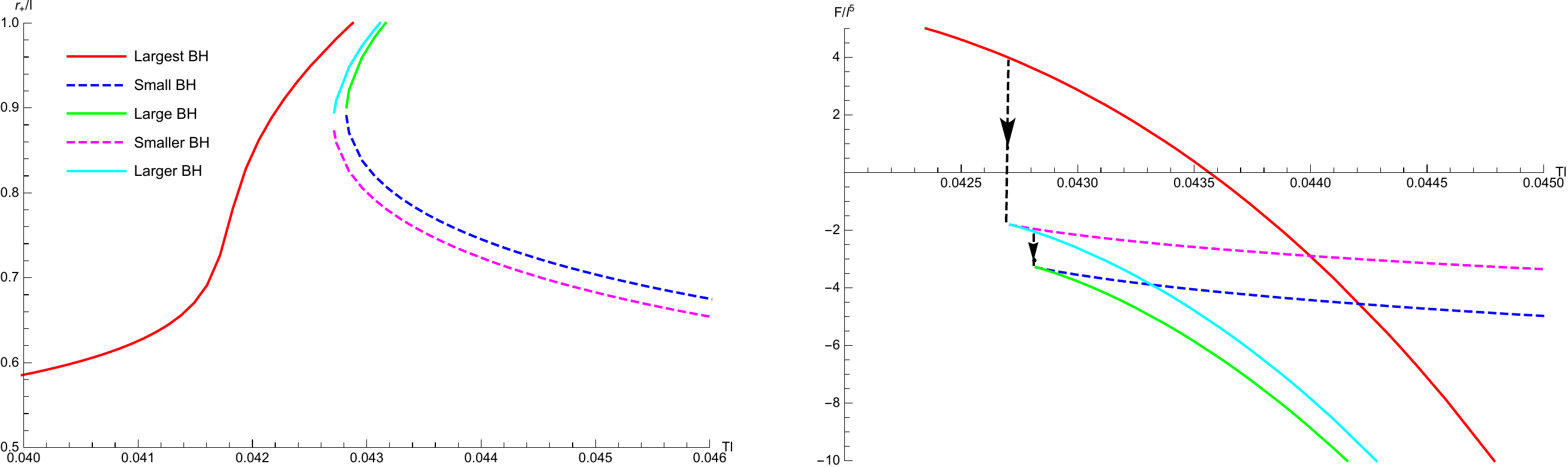}\label{fig:R234:R2}} 
\subfigure[{~ \scriptsize Region III: $h/l^{1.5}=0.5$ and $q/l^{4.5}=0.1$. There is a first order phase transition between Smallest BH and Largest BH.}]{
\includegraphics[width=1\textwidth]{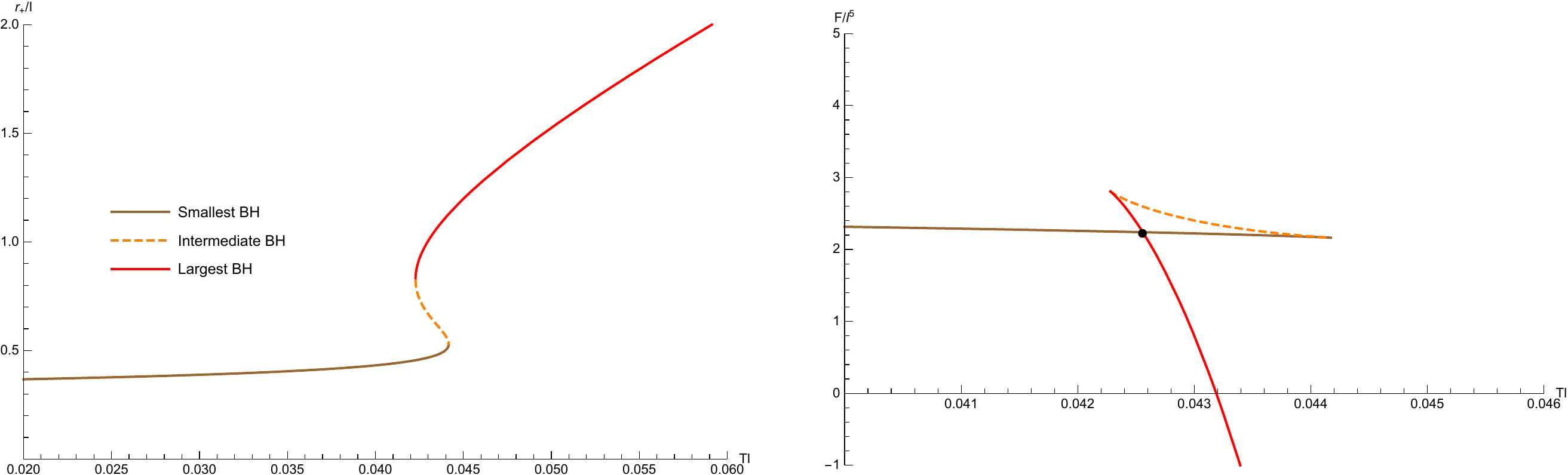}\label{fig:R234:R3}} 
\subfigure[{~ \scriptsize Region IV:$h/l^{1.5}=0.5$ and $q/l^{4.5}=0.3$. There is no phase transition.}]{
\includegraphics[width=1\textwidth]{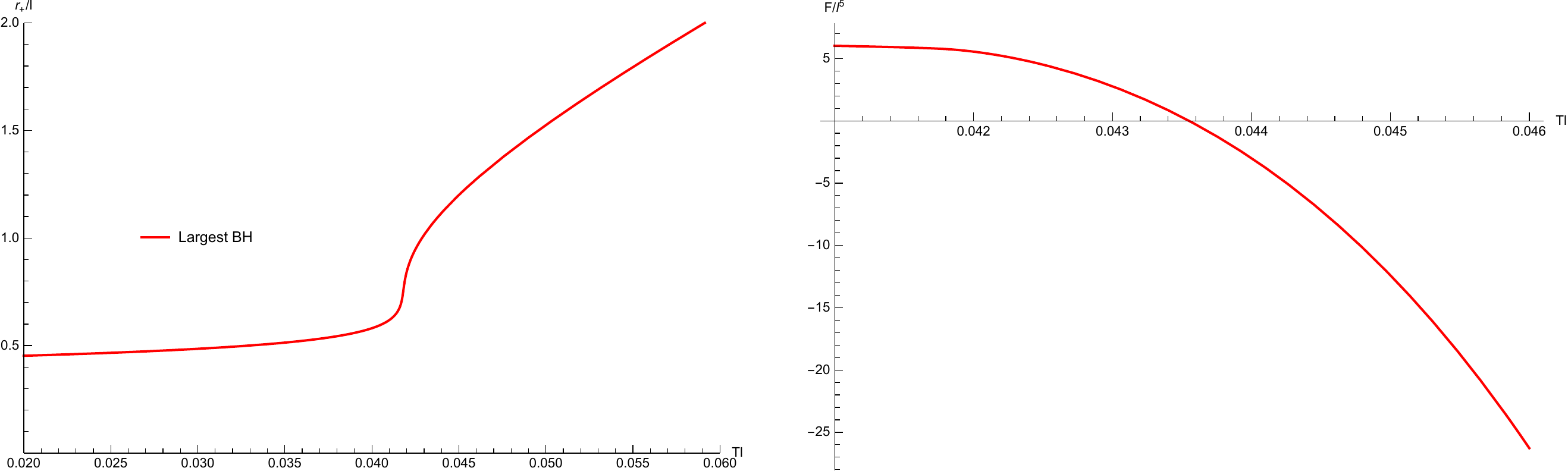}\label{fig:R234:R4}}
\end{center}
\caption{{\protect\footnotesize Plot of $\tilde{r}_{+}$, $\tilde{F}$ against 
$\tilde{T}$ for PL-AdS black holes in RegionsII, III and IV. The number of
branchs is different in these regions. The intermediate BH, small BH and
smaller are always thermally unstable, others are always thermally stable.}}
\label{fig:R234}
\end{figure}

\begin{itemize}
\item Region II: $x<1$ and $\tilde{T}_{1}^{\prime}\left( \tilde{r}%
_{1}\right) \geq0$. In this region, $\tilde{T}_{1}^{\prime}\left( \tilde {r}%
_{+}\right) \geq\tilde{T}_{1}^{\prime}\left( \tilde{r}_{1}\right) \geq0$ and
hence $\tilde{T}_{1}\left( \tilde{r}_{+}\right) $ is\ an increasing
function. So there is only one branch for $\tilde{r}_{+1}(\tilde{T})$:
largest BH, which is thermally stable, there are two branches for $\tilde{r}%
_{+2}(\tilde{T})$: small BH for $0\leq\tilde{T}_{2}\leq\tilde{T}_{2\text{min}%
}$ and large BH for $\tilde{T}_{2}\geq$ $\tilde{T}_{2\text{min}}$, and there
are two branches for $\tilde{r}_{+3}(\tilde{T})$: smaller BH for $0\leq 
\tilde{T}_{3}\leq\tilde{T}_{3\text{min}}$ and larger BH for $\tilde{T}%
_{3}\geq$ $\tilde{T}_{3\text{min}}$, and these are displayed in the left
panel of FIG. \ref{fig:R234:R2}. The Gibbs free energy of the three branches
is plotted in the right panel. The large BH, larger BH and largest BH
branches are thermally stable. (If $x=1,$ it is similar except $\tilde{T}%
_{2} $ and $\tilde{T}_{3}$ merge into one.)

\item Region III: $x>1$ and $\tilde{T}_{1}^{\prime}\left( \tilde{r}%
_{1}\right) <0$. In this region, $\tilde{T}_{1}^{\prime}\left( \tilde{r}%
_{+}\right) =0$ has two solutions $\tilde{r}_{+}=$\ $\tilde{r}_{\text{max}}$
and\ $\tilde{r}_{\text{min}}$ with $\tilde{r}_{\text{max}}<\tilde{r}_{1}<%
\tilde{r}_{\text{min}}$. Since $\tilde{T}_{1}\left( +\infty\right) =+\infty$%
, $\tilde{T}_{1}\left( \tilde{r}_{+}\right) $ has a local maximum of $\tilde{%
T}_{1\text{max}}=\tilde{T}_{1}\left( \tilde{r}_{\text{max}}\right) $ at $%
\tilde{r}_{+}=$\ $\tilde{r}_{\text{max}}$ and a local minimum of $\tilde{T}%
_{1\text{min}}=\tilde{T}_{1}\left( \tilde{r}_{\text{min}}\right) $ at $%
\tilde{r}_{+}=$\ $\tilde{r}_{\text{min}}$. There are three branches for $%
\tilde{r}_{+1}(\tilde{T})$: smallest BH for $0\leq\tilde{T}_{1}\leq\tilde{T}%
_{1\text{max}}$, intermediate BH for $\tilde{T}_{1\text{min}}\leq\tilde{T}%
_{1}\leq\tilde{T}_{1\text{max}}$ and largest BH for $\tilde {T}_{1}\geq$ $%
\tilde{T}_{1\text{min}}$. $\tilde{r}_{+2}(\tilde{T})$ and $\tilde{r}_{+3}(%
\tilde{T})$ don't exist, so there is a first order phase transition. These
are displayed in the left panel of FIG. \ref{fig:R234:R3}. The Gibbs free
energy of the three branches is plotted in the right panel. The smallest BH
and largest BH branches are thermally stable.

\item Region IV: $x>1$ and $\tilde{T}_{1}^{\prime}\left( \tilde{r}%
_{1}\right) \geq0$. In this region, $\tilde{T}_{1}^{\prime}\left( \tilde {r}%
_{+}\right) \geq\tilde{T}_{1}^{\prime}\left( \tilde{r}_{1}\right) \geq0$ and
hence $\tilde{T}_{1}\left( \tilde{r}_{+}\right) $ is\ an increasing
function. So there is only one branch for $\tilde{r}_{+1}(\tilde{T})$:
largest BH, which is thermally stable, $\tilde{r}_{+2}(\tilde{T})$ and $%
\tilde{r}_{+3}(\tilde{T})$ don't exist, so there is no phase transition in
this region. These are displayed in the left panel of FIG. \ref{fig:R234:R4}%
. The Gibbs free energy is plotted in the right panel.
\end{itemize}

\begin{figure}[ptb]
\begin{center}
\subfigure[{~ \scriptsize The four main regions in the $\tilde{h}$-$\tilde{q}$ plane, each of which possesses the
distinct behavior of the branches and the phase structure.}]{
\includegraphics[width=0.48\textwidth]{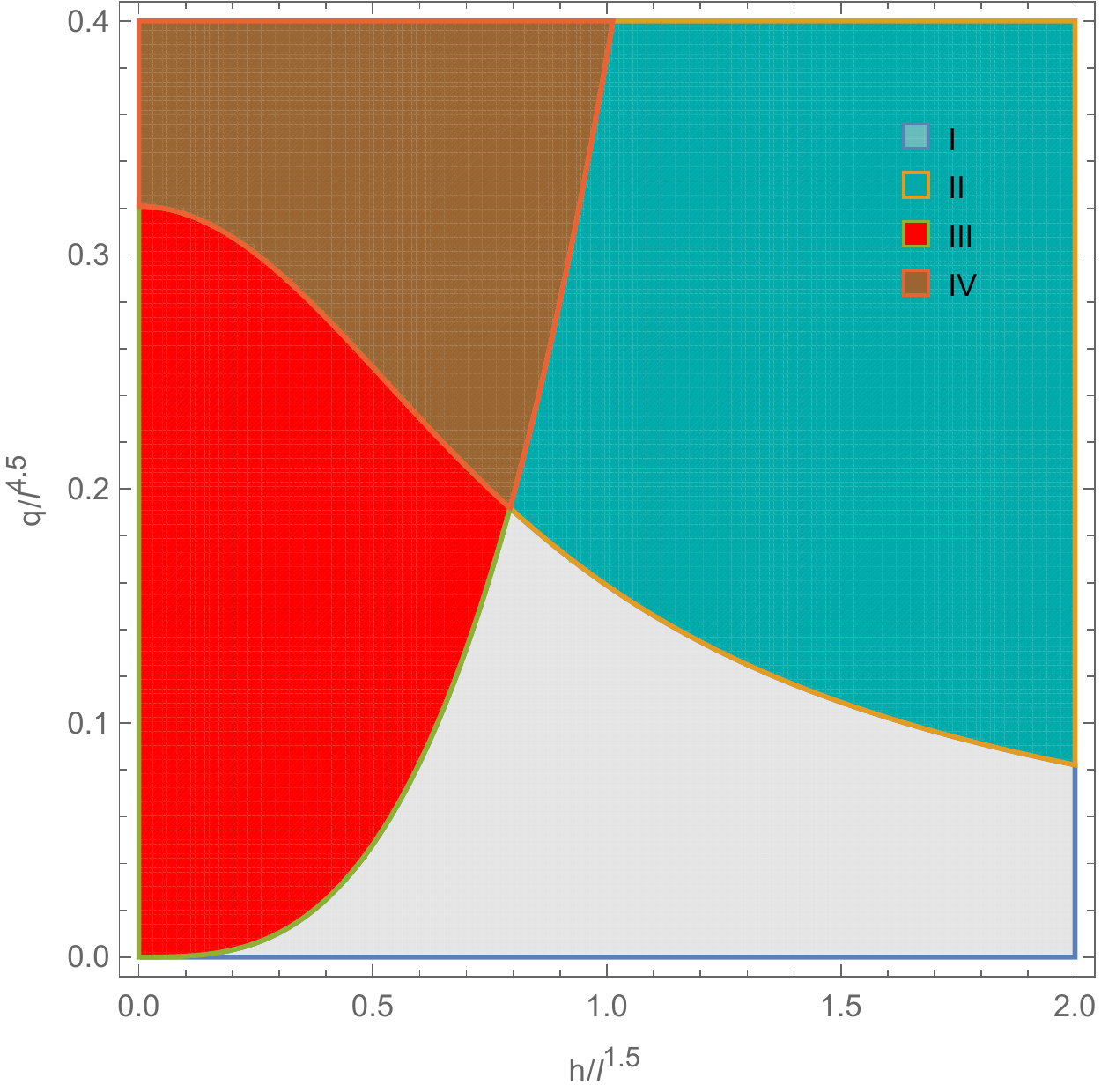}\label{fig:Re:a}} 
\subfigure[{~ \scriptsize  The five subregions of Region I in the $\tilde{h}$-$\tilde{q}$ plane, each of which possesses the
distinct phase structure.}]{
\includegraphics[width=0.48\textwidth]{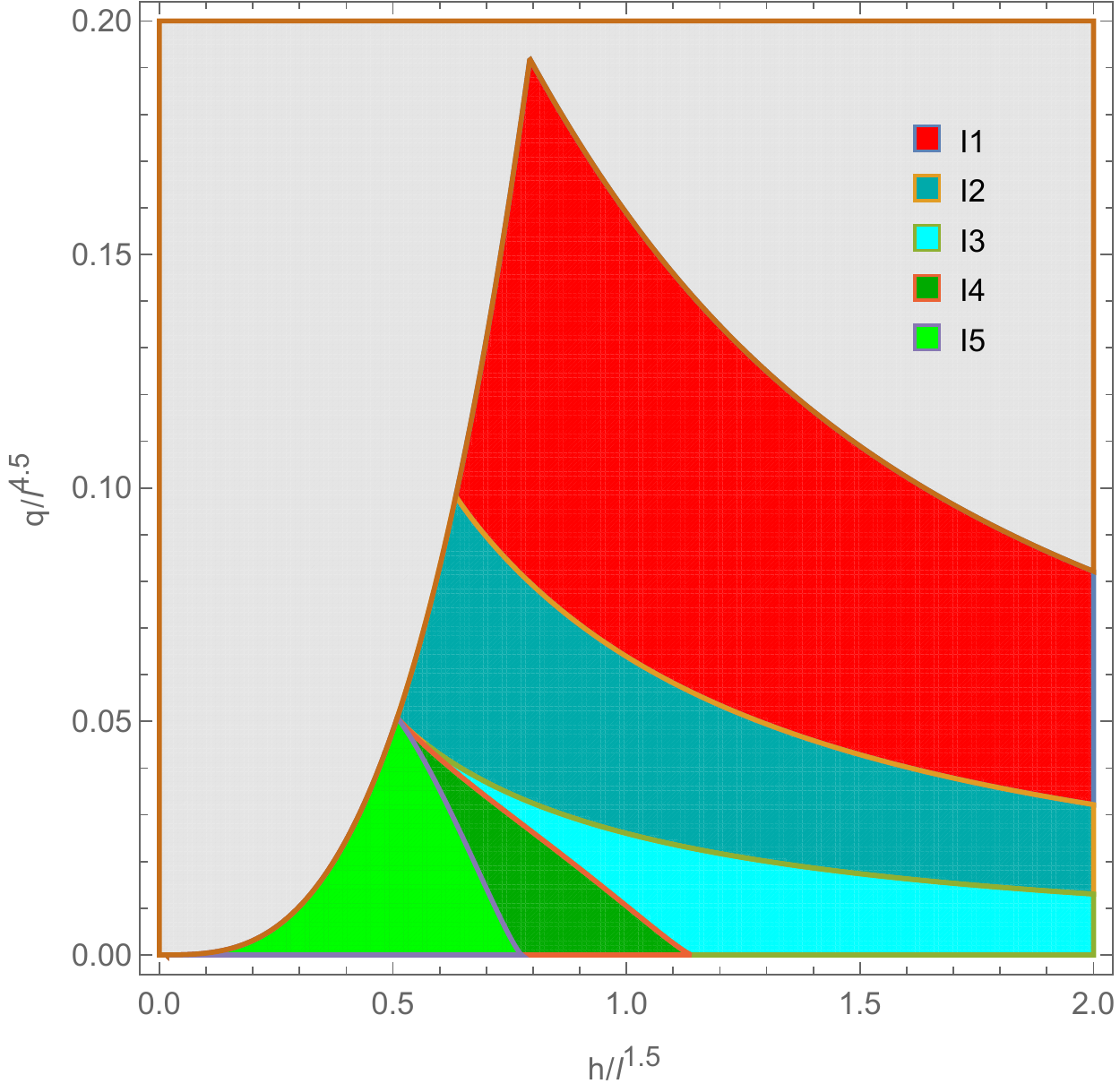}\label{fig:Re:b}}
\end{center}
\caption{{\protect\footnotesize The eight regions in the $\tilde{h}$-$\tilde{%
q}$ plane for dyonic PMI AdS black holes. The color represents for the
branch which it enters at first phase transition with increasing $\tilde{T}$
from 0 (except Region IV), the darker color represents for zeroth order
phase transition, the lighter color represents for first order phase
transition.}}
\label{fig:Re}
\end{figure}

We have marked first order phase transition with black point and zeroth
order phase transition with arrow. In FIG. \ref{fig:Re}, we plot these eight
regions in the $\tilde{h}$-$\tilde{q}$ plane. 

\begin{figure}[ptb]
\begin{center}
\includegraphics[width=0.48\textwidth]{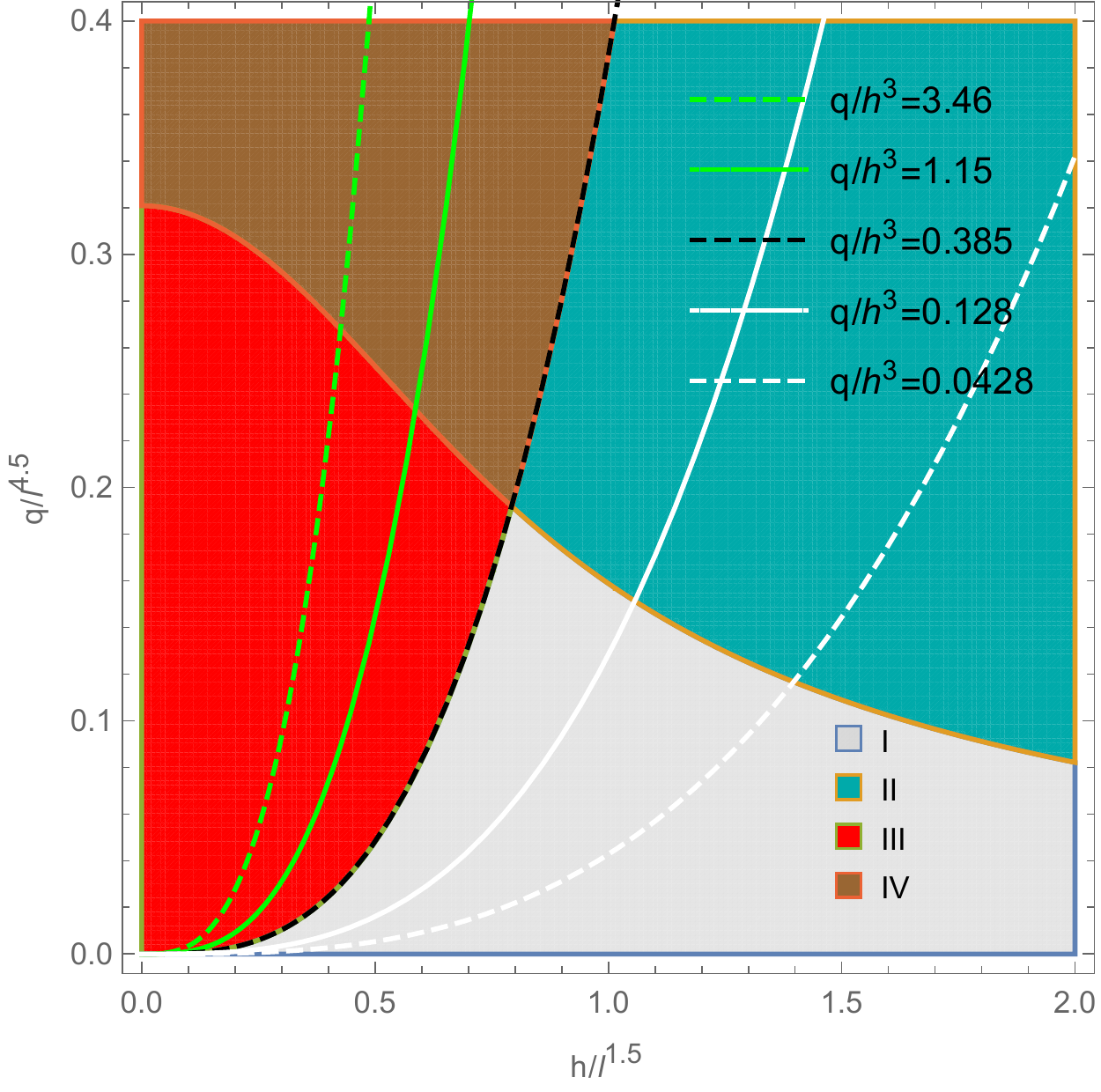}\label{fig:fixp:a} %
\includegraphics[width=0.48\textwidth]{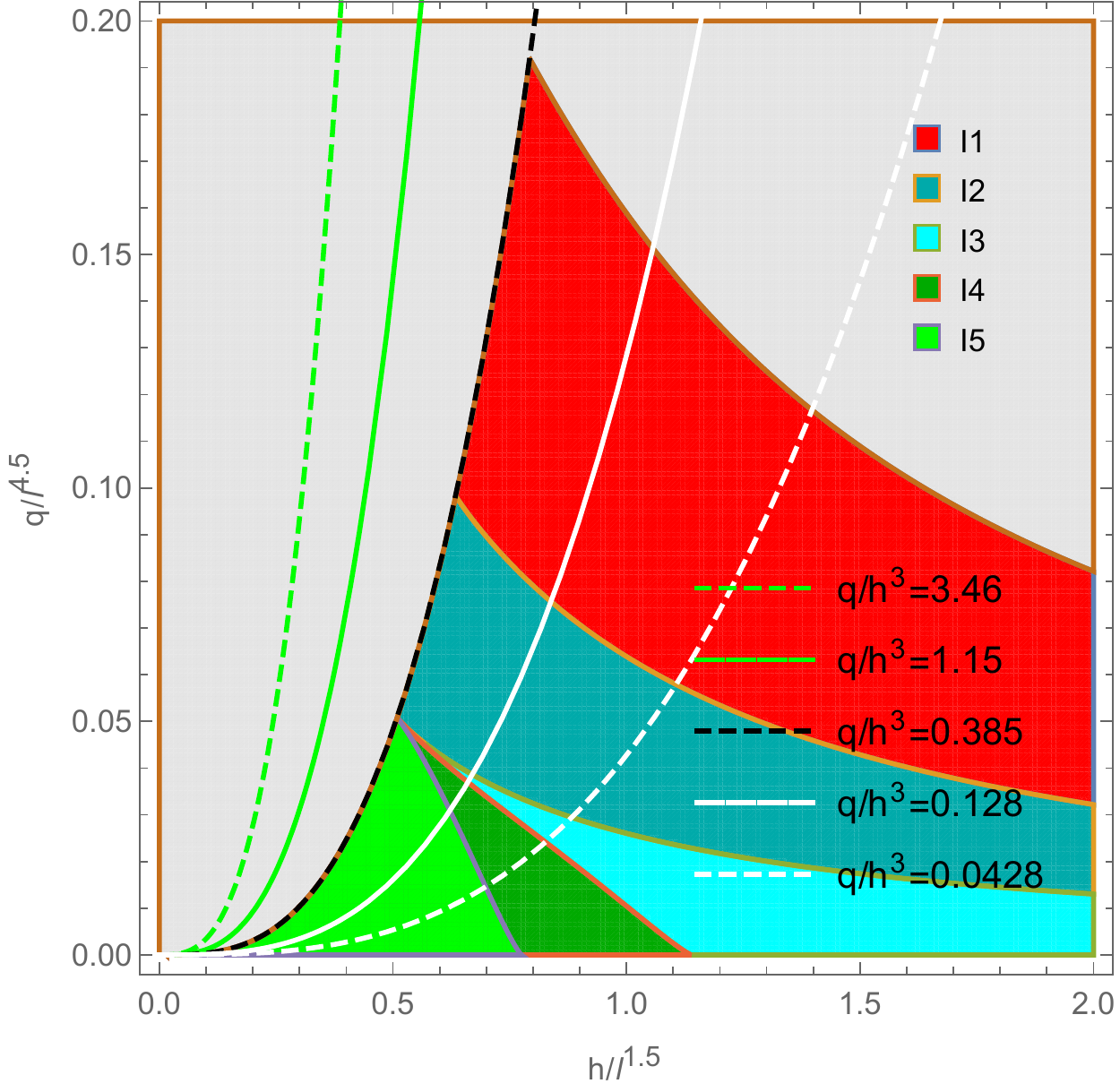}\label{fig:fixp:b}
\end{center}
\caption{{\protect\footnotesize In the case of varying $P$ with fixed $q$
and $h$, the system moves along $\tilde{q}_{l}\left( \tilde{h}\right) $,
which is displayed for various values of $q/h^{3}$. There is always a
critical point and the corresponding largest BH/smallest BH first order
phase transition. }}
\label{fig:fixp}
\end{figure}

We can now discuss the critical behavior and phase structure of black holes
in two cases. The critical line is the boundary between the region in which $%
\tilde{T}\left( \tilde{r}_{+}\right) $ has $n$ extremums and that in which $%
\tilde{T}\left( \tilde{r}_{+}\right) $ has $n+2$ extremums, determined by%
\begin{equation}
\frac{\partial\tilde{T}(\tilde{r}_{+},\tilde{Q},\tilde{a})}{\partial\tilde {r%
}_{+}}=0\text{ and }\frac{\partial^{2}\tilde{T}(\tilde{r}_{+},\tilde {Q},%
\tilde{a})}{\partial\tilde{r}_{+}^{2}}=0.  \label{eq:DBIcritical}
\end{equation}

In the first case, $q$ and $h$ are fixed parameters, and the AdS radius $l$
(the pressure $P$) varies. With fixed values of $q$ and $h$, varying $l$
would generate a curve in the $\tilde{h}$-$\tilde{q}$ plane, which is
determined by%
\begin{equation}
\tilde{q}_{l}\left( \tilde{h}\right) =\frac{q}{h^{3}}\tilde{h}^{3}.
\end{equation}
In FIG. \ref{fig:fixp}, we plot $\tilde{q}_{l}\left( \tilde{h}\right) $ for
various values of $q/h^{3}$. It shows that, there is always critical point
for black holes. For $q/h^{3}>2\sqrt{3}/9$, as one starts from $P=0$, $%
\tilde {q}_{l}\left( \tilde{h}\right) $ always crosses the critical line and
enters Region IV, in which there is no phase transition, from Region III, in
which there is a first phase transition between smallest BH and largest BH.
For $q/h^{3}<2\sqrt{3}/9$, as one starts from $P=0$, $\tilde{q}_{l}\left( 
\tilde{h}\right) $ always passes through five subregions, then crosses the
critical line and enters Region II, in which there are two zeroth phase
transitions, from largest BH to larger BH and from larger BH to large BH. 
\begin{figure}[ptb]
\begin{center}
\subfigure[{~\scriptsize $h/l^{1.5}=0.5$}]{
\includegraphics[width=0.48\textwidth]{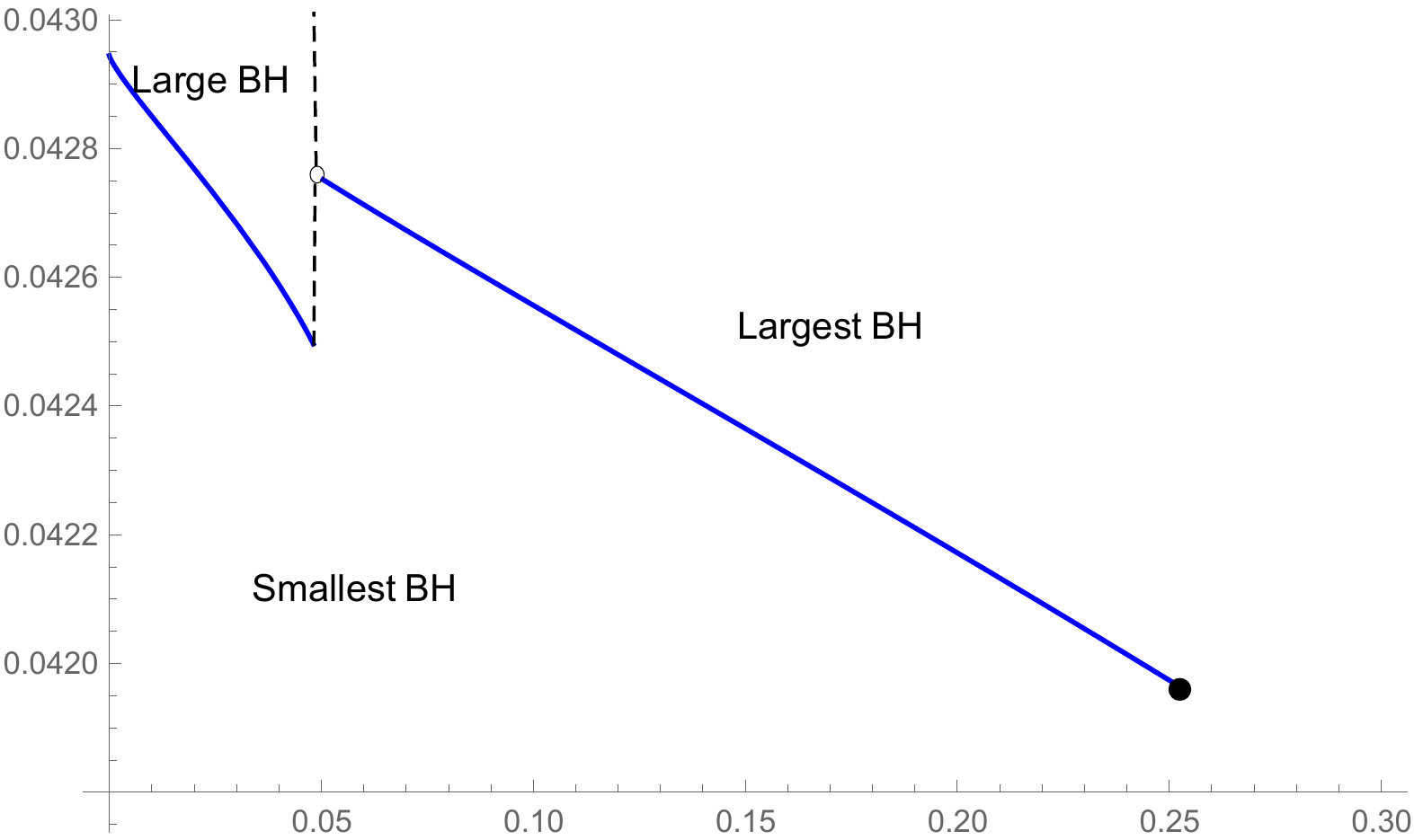}\label{fig:fixh:a}} 
\subfigure[{~\scriptsize $h/l^{1.5}=0.6$}]{
\includegraphics[width=0.48\textwidth]{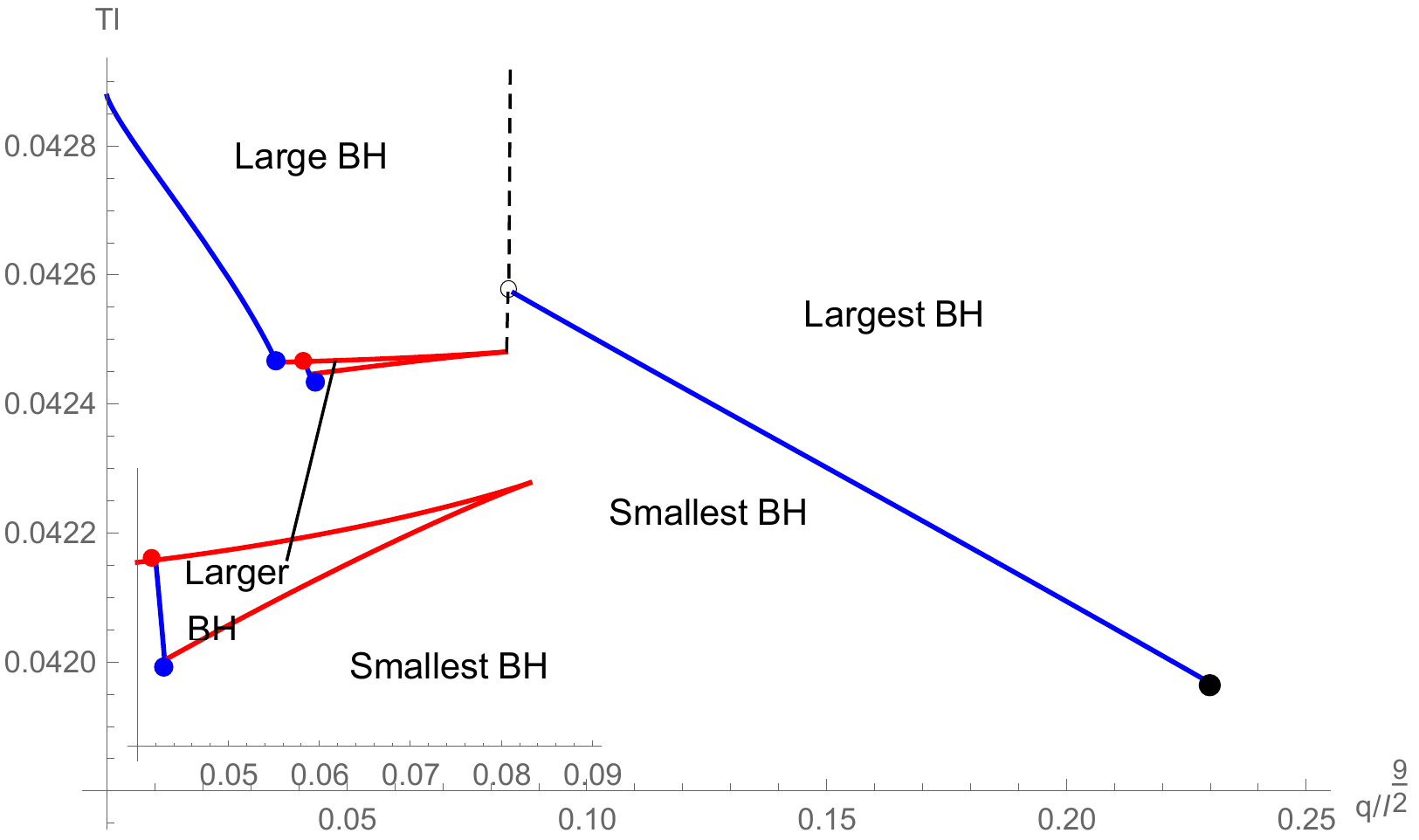}\label{fig:fixh:b}} 
\subfigure[{~\scriptsize $h/l^{1.5}=0.7$}]{
\includegraphics[width=0.48\textwidth]{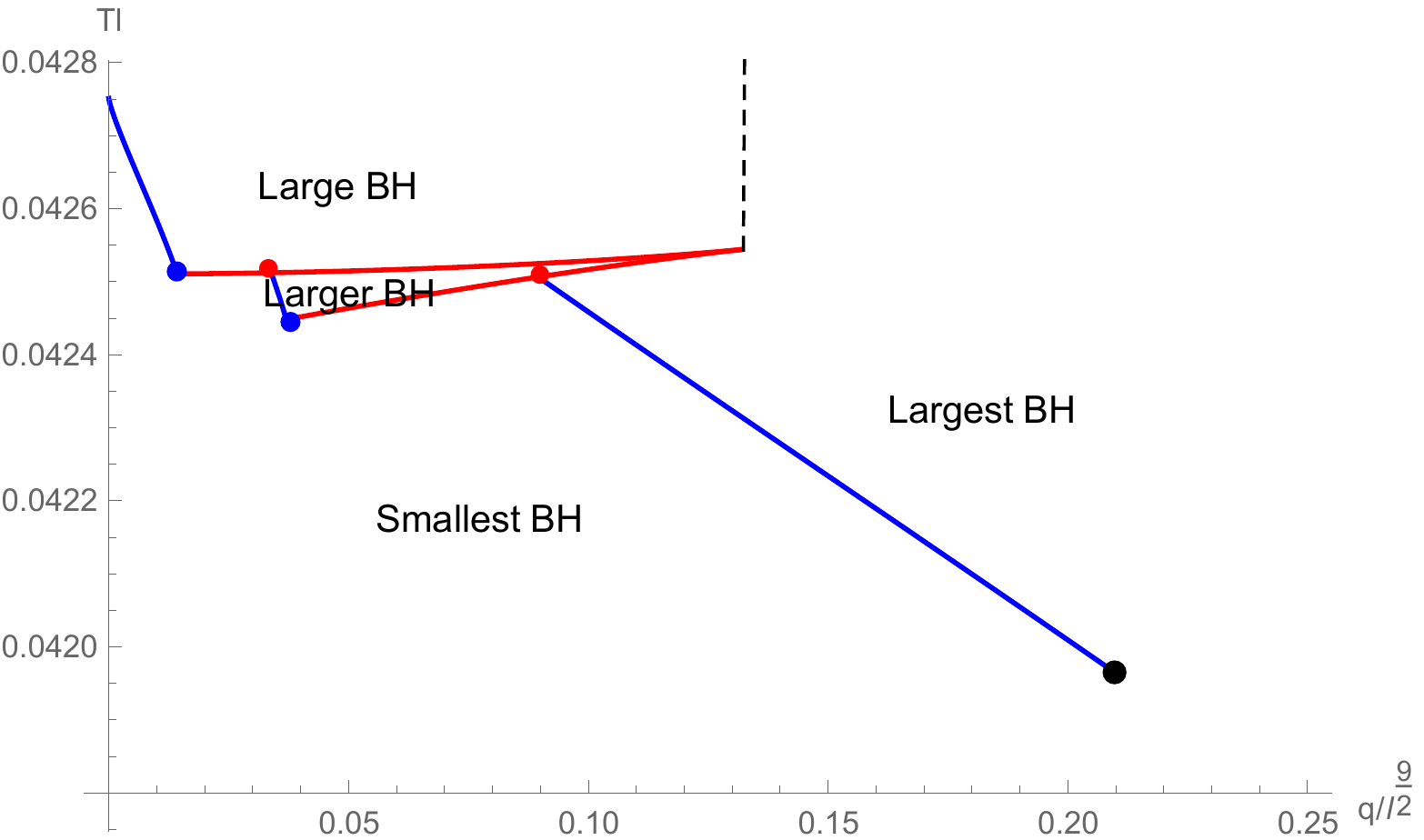}\label{fig:fixh:c}} 
\subfigure[{~\scriptsize $h/l^{1.5}=1$}]{
\includegraphics[width=0.48\textwidth]{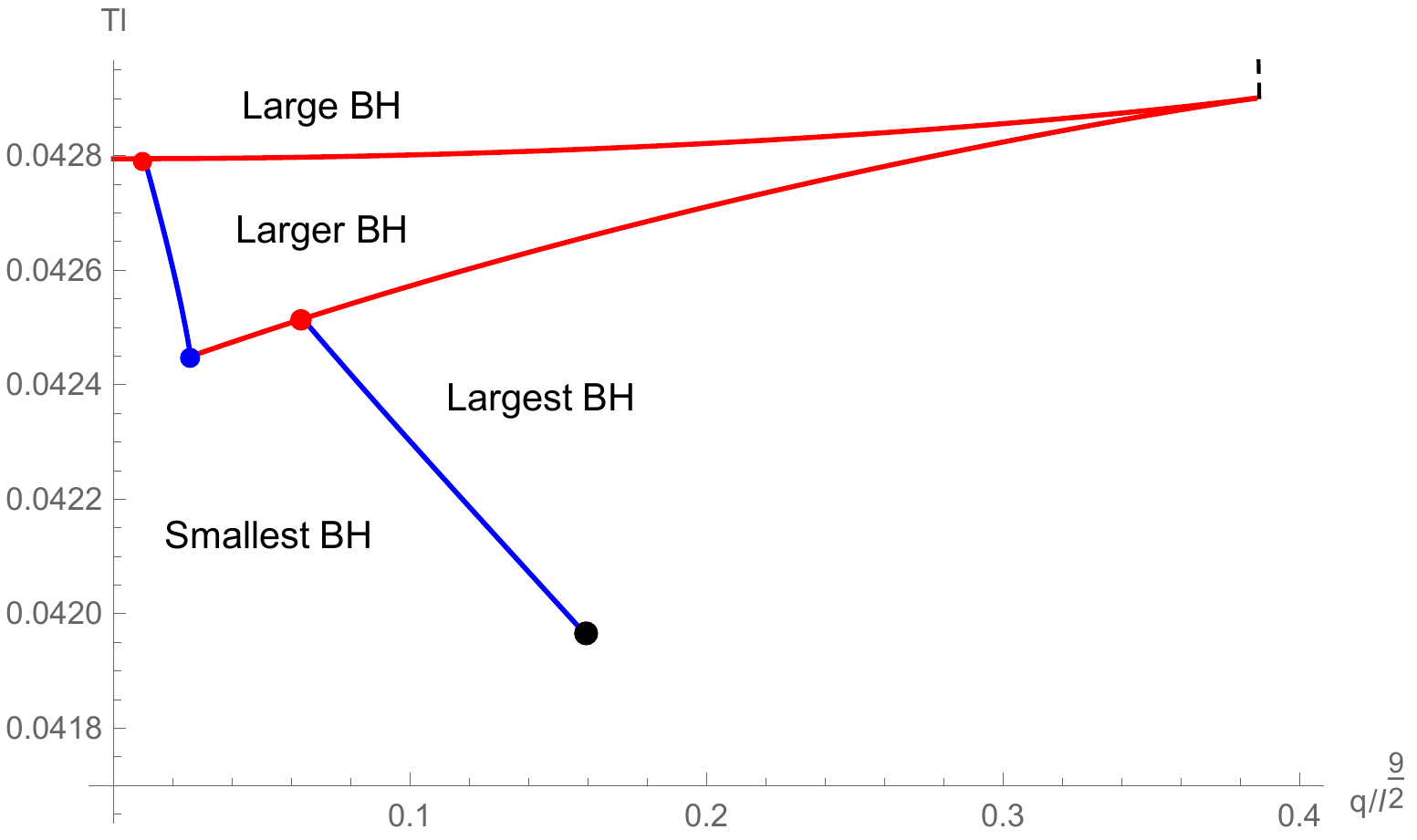}\label{fig:fixh:d}}
\end{center}
\caption{{\protect\footnotesize The phase diagram in the $\tilde{q}$-$\tilde{%
T}$ plane for PMI-AdS black holes with $h/l^{1.5}=0.5$, $h/l^{1.5}=0.6$, $%
h/l^{1.5}=$$0.7$ and $h/l^{1.5}=$$1.$ The Blue line}{\protect\small (point)}%
{\protect\footnotesize represent for first order phase transition and the
Red line {\protect\small (point)} represent for zeroth order transition. The
Black point represent for the critical point.}}
\label{fig:fixh}
\end{figure}

In the second case, $h$ and $P$ $\left( l\right) $ are fixed parameters, and
one varies $\tilde{q}$. As one increases $\tilde{q}$ from $\tilde{q}=0$, the
black hole would experience different regions. And there is always a
critical point for any $\tilde{h}$ since the critical line only has one end
point at $\tilde{h}=0$. In FIG \ref{fig:fixh}, I plot that in the case of $%
\tilde {h}=0.5$, $\tilde{h}=0.6$, $\tilde{h}=0.7$ and $\tilde{h}=1$.

For $\tilde{h}\leq\tilde{h}_{1}\simeq0.51$, as one increases $\tilde{q}$
from $\tilde{q}=0$, the black hole would experience three regions, in which
there occur the smallest BH/large BH first order phase transition $%
\rightarrow$the smallest BH/largest BH first order phase transition $%
\rightarrow$ no phase transition, showed in FIG. \ref{fig:fixh:a}.

For $\tilde{h}\geq\tilde{h}_{1}$, there is another region to across between
the two regions mentioned above, in which there occurs the smallest BH/large
BH zeroth order phase transition $\rightarrow$a smallest BH/larger BH first
order phase transition and a larger BH/large BH zeroth phase transition$%
\rightarrow$ a smallest BH/larger BH zeroth order phase transition and a
larger BH/large BH zeroth phase transition. These are showed in FIG. \ref%
{fig:fixh:b}.

When $\tilde{h}\geq\tilde{h}_{2}\simeq0.63$, the range of region that has a
smallest BH/largest BH first order phase transition will expand to the left,
which is showed in FIG. \ref{fig:fixh:c}.

When $\tilde{h}\geq\tilde{h}_{3}\simeq0.76$, the smallest BH/large BH first
order phase transition at left disappears, showed in FIG. \ref{fig:fixh:a}
and the smallest BH/large BH zeroth order phase transition disappears when $%
\tilde{h}\geq\tilde{h}_{4}\simeq1.14$, it is reminiscent of FIG. \ref%
{fig:fixh:d}, just cutting the Red line before Blue line at left.

\section{Discussion and Conclusion}

\label{Sec:Con}

We have investigated the thermodynamic behavior of $d$-dimensional dyonic PM
AdS black holes in an extended phase space, which includes the conjugate
pressure/volume quantities. It showed that the black hole's temperature $T$,
charge $q$, horizon radius $r_{+}$ (thermodynamic volume $V$), the AdS
radius $l$ (pressure $P$) and the magnetic parameter $h$ could be connected
by 
\begin{equation}
Tl=\tilde{T}\left( r_{+}/l,q/l^{a},h/l^{b}\right) ,
\end{equation}
where $a$ and $b$ depend on the dimension $d$ and the power exponent $p$. In
the canonical ensemble with fixed $T$ and $q$, we found that the critical
behavior and phase structure of the black hole are determined by $\tilde {q}%
\equiv q/l^{a}$ and $\tilde{h}\equiv h/l^{b}$.

For 8-dimensional PM dyonic AdS black holes with an power exponent of $2$,
we examined their critical behavior and phase structure, whose dependence on 
$\tilde{q}$ and $\tilde{h}$ was plotted in FIG. \ref{fig:Re}. There are 8
regions in FIG. \ref{fig:Re}, and each region has a different phase
behavior. Unlike other black holes, the temperature $\tilde{T}\left(
r_{+}/l,q/l^{4.5},h/l^{1.5}\right) $ of PM dyonic AdS black holes could be
more than one when some parameter configurations of $q/l^{4.5}$ and $%
h/l^{1.5}$ vary. If we discuss them separately, one of them ($\tilde{T}_{1}$
in this case) likes the case of RN-AdS black holes and the other two (if
exist) like the case of Schwarzschid-AdS black holes. The combination of
them results in these rich phase structures and phase behaviors.

The thermodynamically preferred phases, along with the zeroth and first
order phase transitions and critical points, were displayed in FIG. \ref%
{fig:fixh} for the black holes. We examined thermal stabilities of the black
holes and found that all the thermodynamically preferred phases are
thermally stable.

\section{Acknowledgement}

We are grateful to thank Peng Wang and Yuchen Huang for useful discussions.
This work is supported by NSFC (Grant No.11947408).

\bibliographystyle{unsrturl}
\bibliography{TP-PM}

\end{document}